\documentclass[aoas]{imsart}

\RequirePackage{amsthm,amsmath,amsfonts,amssymb}
\RequirePackage[authoryear]{natbib}
\usepackage{natbib,graphicx,lscape,longtable}
\usepackage{hyperref}
\usepackage{bm}
\usepackage{subfig}
\usepackage{xcolor}
\usepackage{colortbl}
\startlocaldefs
\theoremstyle{plain}
\bibpunct{(}{)}{;}{a}{,}{,}

\newtheorem{theorem}{Theorem}
\newtheorem{lemma}{Lemma}
\newtheorem{proposition}{Proposition}

\allowdisplaybreaks[4]

\hypersetup{
	colorlinks=true,
	linkcolor=blue,
	citecolor=blue,
	urlcolor=blue}

\def\beq{\begin{equation}}
	\def\eeq{\end{equation}}
\def\beqr{\begin{eqnarray}}
	\def\eeqr{\end{eqnarray}}
\def\beqrs{\begin{eqnarray*}}
	\def\eeqrs{\end{eqnarray*}}
\def\bet{\begin{theorem}}
	\def\eet{\end{theorem}}
\def\bel{\begin{lemma}}
	\def\eel{\end{lemma}}
\def\bep{\begin{proposition}}
	\def\eep{\end{proposition}}
\def\bg{\begin{figure}[tbph]\begin{center}}
		\def\eg{\end{center}\end{figure}}

\def\bc{\begin{center}}
	\def\ec{\end{center}}

\def\tr{\mbox{tr}}

\def\ve{\varepsilon}

\def\blue{\color{blue}}

\def\wt{\widetilde}
\def\wh{\widehat}

\def\SE{\widehat{\mbox{SE}}}

\def\mA{\mathbb A}

\def\b0{\mathbf{0}}

\def\ve{\varepsilon}

\def\mR{\mathbb{R}}
\def\mS{\mathcal S}

\def\mL{\mathcal L}
\def\btheta{\boldsymbol{\theta}}
\def\bgamma{\boldsymbol{\gamma}}
\def\bW{\boldsymbol{W}}

\def\blue{\color{blue}}

\def\mW{\mathbb{W}}

\def\var{\mbox{var}}
\def\avar{\mbox{avar}}

\def\cov{\mbox{cov}}
\def\acov{\mbox{acov}}
\def\tr{\mbox{tr}}
\def\argmin{\mbox{argmin}}
\def\argmax{\mbox{argmax}}

\def\diag{\mbox{diag}}

\def\MSE{\mbox{MSE}}
\def\SE{\mbox{SE}}
\def\CP{\mbox{CP}}

\newcommand{\balpha}{\boldsymbol{\alpha}}
\newcommand{\bbeta}{\boldsymbol{\beta}}

\def\bX{\mbox{\boldmath $X$}}
\def\bZ{\mbox{\boldmath $Z$}}

\def\bU{\bm U}

\def\wbD{\wt {\mathbb D}}
\def\mI{\mathbb I}
\def\MSE{\mbox{MSE}}

\newcommand{\mE}{{\mathbb E}}

\numberwithin{equation}{section}

\endlocaldefs

\begin{document}

\begin{frontmatter}
\title{Statistical Inference for Regression with Imputed Binary Covariates with Application to Emotion Recognition}
\runtitle{Statistical Inference for Regression with Imputed Binary Covariates }

\begin{aug}
\author[A]{\fnms{Ziqian}~\snm{Lin}},
\author[B]{\fnms{Danyang}~\snm{Huang}\ead[label=e2]{dyhuang89@126.com}}
\author[A]{\fnms{Ziyu}~\snm{Xiong}},
\and
\author[A]{\fnms{Hansheng}~\snm{Wang}},
\address[A]{Guanghua School of Management, Peking University, Beijing, China}

\address[B]{Center for Applied Statistics and School of Statistics, Renmin University of China, Beijing, China\printead[presep={,\ }]{e2}}
\end{aug}

\begin{abstract}

In the flourishing live streaming industry, accurate recognition of streamers' emotions has become a critical research focus, with profound implications for audience engagement and content optimization. However, precise emotion coding typically requires manual annotation by trained experts, making it extremely expensive and time-consuming to obtain complete observational data for large-scale studies. Motivated by this challenge in streamer emotion recognition, we develop here a novel imputation method together with a principled statistical inference procedure for analyzing partially observed binary data. Specifically, we assume for each observation an auxiliary feature vector, which is sufficiently cheap to be fully collected for the whole sample. We next assume a small pilot sample with both the target binary covariates (i.e., the emotion status) and the auxiliary features fully observed, of which the size could be considerably smaller than that of the whole sample. Thereafter, a regression model can be constructed for the target binary covariates and the auxiliary features. This enables us to impute the missing binary features using the fully observed auxiliary features for the entire sample. We establish the associated asymptotic theory for principled statistical inference and present extensive simulation experiments, demonstrating the effectiveness and theoretical soundness of our proposed method. Furthermore, we validate our approach using a comprehensive dataset on emotion recognition in live streaming, demonstrating that our imputation method yields smaller standard errors and is more statistically efficient than using pilot data only. Our findings have significant implications for enhancing user experience and optimizing engagement on streaming platforms.

\end{abstract}


\end{frontmatter}

\section{Introduction}

\begin{figure}[t]
	\centering
	\includegraphics[width = 1.0\columnwidth]{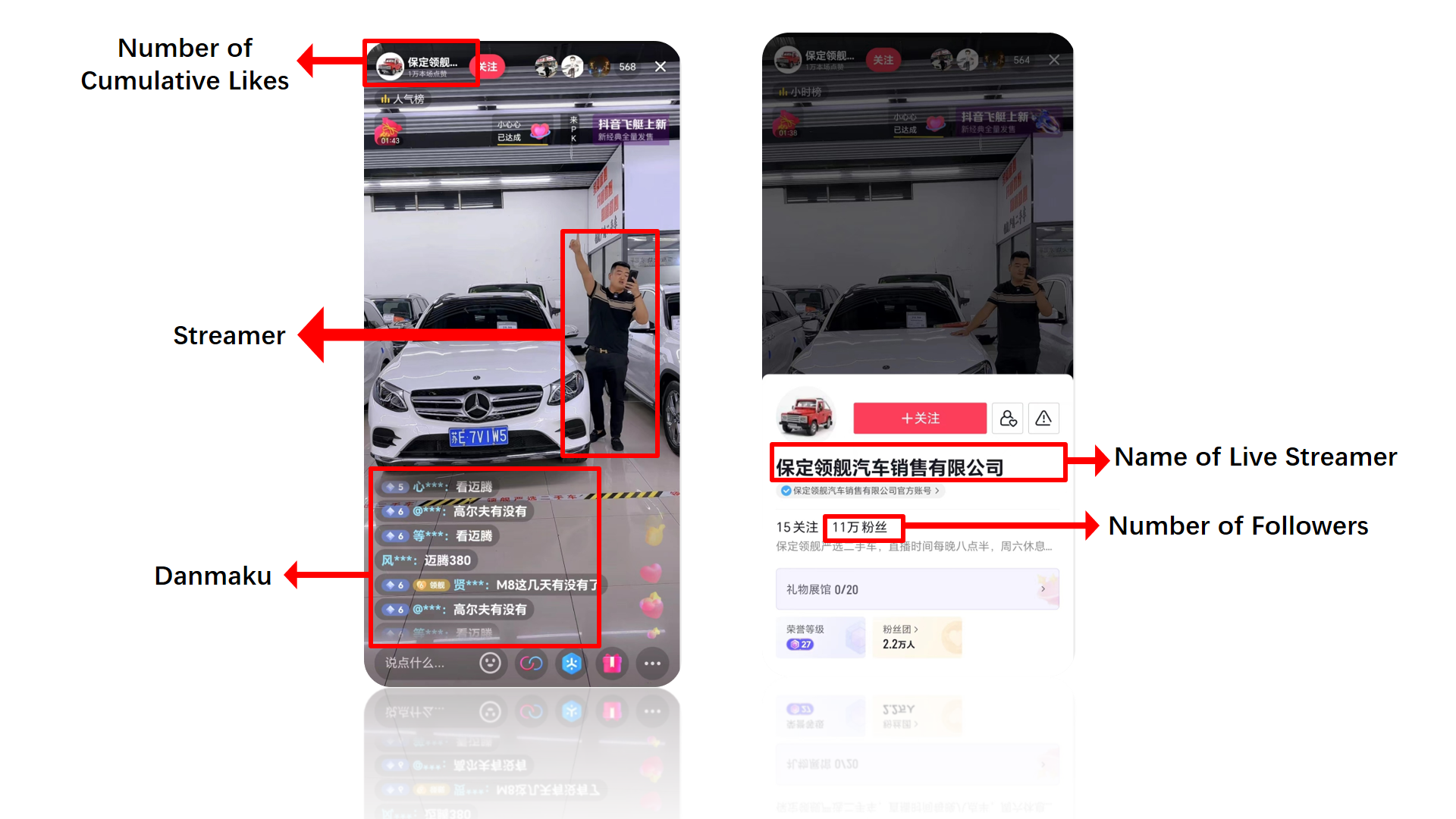}
	\caption{ Example of live streaming on TikTok. The left figure represents the live streaming interface as seen by an application user. The right figure represents the information interface of the streamer. The red box highlights some important components for the live streaming. }
	\label{Figure live streaming}
\end{figure}

In empirical research, researchers often rely on various regression models to study the dependence relationship between a response of interest $Y$ and a set of covariates, which can be further decomposed into two parts. The first part contains a feature vector $\bm X$, which is fully observed. The second part contains another set of features $\bm Z$, which may be incomplete owing to the data collection cost. Here, we are particularly interested in the case where $\bm Z$ is of a binary vector form. In other words, each component of $\bm Z$ takes the value of either 1 or 0, representing the binary status of the subject under the study. Consider the example of live streaming (see Figure \ref{Figure live streaming}). The emotional state of the streamers (i.e., 1 represents a positive emotion and 0 a nonpositive one) is of great importance for promoting the number of new likes (i.e., $Y$), which could be one component of $\bm Z$. Accordingly, we need to record the emotion of the streamer for every sampled moment, which is typically an audio clip lasting 5 seconds. Subsequently, human effort is needed to study the video clips and code the emotional state of the streamer appropriately. In most cases, a live streaming could last for one or two hours. This might lead to a sample with thousands of video clips, whose emotional state must be coded accurately by human efforts. Unfortunately, we have in hand live streaming video records of tens of thousands of hours in length resulting in a huge sample tens of millions in size. If the emotional state of all the sampled moments are to be accurately coded, the human effort required is extremely significant. Hence, how to solve this issue becomes an important problem.

It is remarkable that the coding problem described above is not rare in practical research. For example, \cite{zhang2022makes} studied how the images taken by Airbnb's photographers will affect the property demand. In their research, more than 510,000 images were involved. For each image, a 6-dimensional binary feature vector needed to be coded. The six components of this binary vector were whether the image was of high quality, whether the image belonged to bathrooms, bedrooms, kitchens, living rooms, or outdoor areas. Practically, if human effort was required to code the binary vector for every image, the total cost would be extremely high. As another example, \cite{zhang2023can} studied how the reviews posted on Yelp affect the survival of corresponding restaurants. In their study, a total of 1,121,069 reviews were involved. For each review, the binary vector contained four components. They corresponded to whether food, service, environment, and price were mentioned in the review. If human effort was required to code all the 1,121,069 review samples, the cost would be extremely high. Similar problems are also encountered in \cite{liu2019large} for consumer review research, \cite{timoshenko2019identifying} for marketing research, and \cite{li2023estimating} for house renting research. Given that large-scale and unstructured datasets are becoming increasingly available, similar problems are expected to occur more frequently.

A practical and popular solution is to accurately collect for each observation a feature vector $\bm W$, which should be highly related to $\bm Z$ but incur a very low collection cost \citep{liu2019large,zhang2022makes,zhang2023can}. For example, for our intended application for emotion recognition, this $\bm W$-vector is a feature vector extracted from the audio record of the streamers using the Mel spectrogram and neural networks. As can be seen, this can be automatically generated by the live streaming system and thus is cheap to collect. However, we do believe it to be extremely useful for accurately identifying the binary vector of interest $\bm Z$, as human emotions are largely reflected in their voices. Therefore, we should reasonably expect that the true emotional state $\bm Z$ can be predicted by the audio feature vector $\bm W$ to some extent. Meanwhile, the feature vector $\bm W$ is typically highly abstract and can hardly be interpreted directly. Therefore $\bm W$ cannot be included in the regression model for explaining $Y$ directly. Otherwise, the interpretability of the whole model can be largely affected. Then, a natural question is: How do we construct a statistical model from $\bm W$ to $\bm Z$? For convenience, we use $\wh {\bm Z}$ to represent this predicted $\bm Z$ value (e.g., estimated emotional state). Subsequently, one might consider treating $\wh {\bm Z}$ as if it were the true $\bm Z$. Following the past literature, we refer to this operation as imputation \citep{little2019statistical}. Thereafter, a standard regression model can be constructed. The parameter of interest can be estimated and the hypotheses can be tested.

However, an important problem remains unsolved: building a model from $\bm W$ to $\bm Z$. To this end, a subsample with accurately collected $\bm W$ and $\bm Z$ is inevitably needed. For convenience, we refer to this subsample as the \textit{pilot sample} \citep{pan2022note}. As mentioned before, collecting information about $\bm Z$ is practically expensive because human effort is necessary. Therefore, the size of this pilot sample is unlikely to be large. For example, the size of the pilot sample used in \cite{zhang2022makes} is 2,259, which accounts for approximately 0.4\% of the total sample size. Similarly, the size of the pilot sample in \cite{zhang2023can} is 10,000, which accounts for approximately 0.9\% of the total sample. With a limited budget, one often wishes the size of this pilot sample to be as small as possible. However, it cannot be too small, otherwise the imputed binary covariates $\wh {\bm Z}$ could be considerably different from the true ones, which in turn makes the subsequent statistical estimation inaccurate. Once all the missing covariates $\bm Z$ are imputed by $\wh {\bm Z}$, a standard statistical estimation procedure (e.g., maximum likelihood estimation) can be applied by treating the imputed covariates $\wh {\bm Z}$ as if they were the true $\bm Z$. This immediately leads to an important question: What is the asymptotic behavior of the resulting practical estimator? 

Note that the problem we studied here is closely related to the imputation literature. In fact, regression problems with partially imputed data have been studied extensively in the past literature \citep{little2019statistical}. There are two streams in the imputation literature. The first stream focuses on the imputation of a missing response. For example, \cite{shao2002sample} studied imputing the nonresponses in survey data by a joint regression model to compute the sample correlation coefficients. \cite{qin2008efficient} proposed an imputation method robust to misspecification in the missing mechanism or parametric regression models. \cite{fang2009pseudo} studied the asymptotic properties of the mean estimators in survey data when the nonrespondents are imputed based on pseudo empirical likelihood estimators. 
It is remarkable that, for all these pioneer studies, all the feature vectors (i.e., $\bm X,\bm Z$ and $\bm W$) should be included in the regression model for $Y$. In our case, this means that the feature vector $\bm W$ is to be included in the regression model for $Y$. As explained before, in our case $\bm W$ is a very indirect and abstract vector extracted from unstructured data (e.g., video and audio). Therefore $\bm W$ can hardly be interpreted directly and thus cannot be included in the main regression model.
The second stream of literature focuses on imputation of covariates. For example, \cite{white2009imputing} considered the imputation of the normal and binary covariates for the Cox proportional hazard model. Interesting imputation algorithms were developed but the theoretical properties of the estimators, such as consistency and asymptotic normality, were not carefully discussed. Meanwhile, various multiple imputation methods were developed by for example \cite{wei2012multiple} for quantile response and \cite{wang2012multiple} for M-estimation. The key idea of multiple imputation is to impute a missing covariate (e.g., $\bm Z$ in our problem) multiple times according to the estimated conditional distribution of the observed response ($Y$) and covariates ($\bm X$). It is remarkable that the aforementioned literature mainly focuses on imputation for continuous variables.

To summarize, the literature on binary covariate imputation by auxiliary information $W$ is relatively limited. Furthermore, for our analysis task in this study, there are two specific observations through data: (1) positive emotion accounts for only a small fraction in the pilot sample; and (2) deep neural networks have successful predictability in classification tasks \citep{dosovitskiy2020image,hassani2021escaping}. Based on these observations, we studied here two interesting and important special cases, which have never been studied to the best of our knowledge. The first is the \textit{highly imbalanced case}, where the percentage of positive responses is small. Theoretically, we can assume that the positive response probability converges toward 0 as the sample size diverges to infinity. This leads to a highly imbalanced distribution for positive and negative cases for $\bm Z$. The second is the \textit{highly predictable case}, where the regressor of $\bm Z$ can be predicted by $\bm W$ with high accuracy. Theoretically, this amounts to assuming that the conditional variance of $\bm Z$ given $\bm W$ converges toward 0 as the sample size diverges. Both special cases are well motivated by real applications and lead to interesting new findings for asymptotic theory. This makes the study's contribution unique.

\begin{figure}[t]
	\centering
	\includegraphics[width = 0.7\columnwidth]{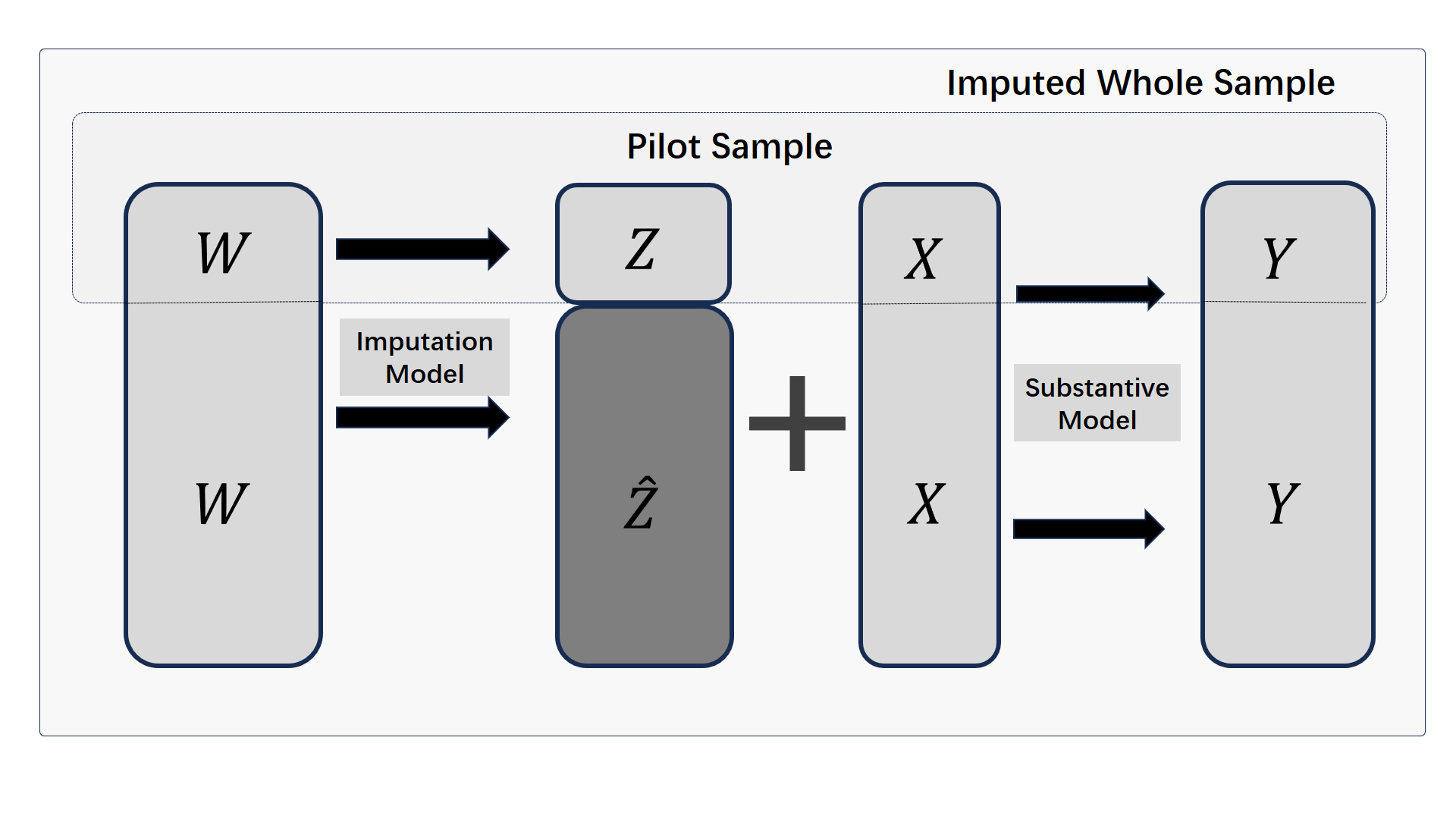}
	\caption{ Framework of regression with
		imputed binary covariates. The pilot sample contains the observations with true $\bm Z$. The imputed whole sample contains all observations with true $\bm Z$ and imputed $\wh {\bm Z}$. }
	\label{fig:frame}
\end{figure}

As our first attempt, we consider here a standard linear regression model for the regression relationship between $Y$ and $(\bm X,\bm Z)$, which is referred to as a substantive model \citep{white2010bias,bartlett2015multiple}. Next, a logistic regression model is considered for each component of $\bm Z$ and the feature vector $\bm W$, which is referred to as an imputation model \citep{white2010bias, bartlett2015multiple}. Therefore, a set of logistic regression models can be trained between $\bm Z$ and $\bm W$ on the pilot sample. Subsequently, the unobserved $\bm Z$ values can be imputed by $\wh {\bm Z}$ for the whole sample. Thereafter, we treat the predicted $\wh {\bm Z}$ as if they were the true $\bm Z$. Then, a standard ordinary least squares (OLS) estimator can be obtained for the whole sample. This leads to the final estimator for the regression coefficient of interest. The framework of the whole procedure is shown in Figure \ref{fig:frame}. 
We show theoretically that the imputation based estimator is consistent and asymptotically normal. Subsequently, asymptotically valid statistical inferences (e.g., confidence intervals) can be constructed.
In particular, the asymptotic theory of the two special cases (i.e., the highly imbalanced and highly predictable cases) are studied with interesting theoretical findings. Extensive simulation experiments are presented to demonstrate the finite sample performance of the proposed method. A real data example of live streaming is presented for illustration purposes.

The remainder of this article is organized as follows. Section 2 develops the methodology with rigorous asymptotic support. Extensive numerical studies are presented in Section 3 to demonstrate the proposed method. Lastly, we conclude this article with a brief discussion in Section 4.

\section{Statistical Inference for Imputed Estimator}
\subsection{The Model Setup}

Let $(\bm Z_i,\bm X_i,Y_i)$ be the $i$-th ($1\le i\le N$) observation independently collected from the joint distribution of $(\bm Z,\bm X,Y)$. Here $Y_i\in\mR$ is the response, $\bm Z_i = (Z_{i1},\dots, Z_{ip})^\top\in\{0,1\}^p$ is the vector of interested binary features, and $\bm X_i = (X_{i1},\dots,X_{iq})^\top \in \mR^q$ is the vector of the fully observed control variables. To model the relationship between $Y_i$ and $(\bm Z_i,\bm X_i)$, we assume the following  correctly specified linear regression model
\beq
Y_i = \bm Z_i^\top \bm \beta + \bm X_i^\top \bm \gamma + \ve_i, \label{linear regression}
\eeq
where $\bm \beta=(\beta_1,\dots,\beta_p)^\top \in \mR^p $ is the interested coefficient vector, $\bm \gamma = (\gamma_1,\dots,\gamma_q)^\top \in \mR^q$ is the coefficient vector of the control variables, and $\ve_i\in \mR$ is the independent random noise satisfying $\mE(\ve_i) = 0$ and $\var(\ve_i)= \sigma^2$. For convenience, we refer to this model as a \textit{substantive model}. To estimate the unknown coefficients $\bm \beta$ and $\bm \gamma$, a standard OLS approach can be applied. Specifically, we minimize the least squares loss function
\[
\mL_{y,\text{lse}} ( \bm \beta,\bm \gamma) = \sum_{i=1}^{N}\Big(Y_i - \bm Z_i^\top  \bm \beta - \bm X_i^\top \bm \gamma\Big)^2.
\]
Then the OLS estimator is given as $(\wh {\bm \beta}_{\text{ols}}, \wh {\bm \gamma}_{\text{ols}}) = \argmin_{\bm \beta,\bm \gamma}\mL_{y,\text{lse}}(\bm \beta,\bm \gamma)$, where $\wh{\bm \beta}_{\text{ols}} = (\wh \beta_{\text{ols},1},\dots,\wh\beta_{\text{ols},p})^\top\in\mR^p$ and $\wh{\bm\gamma}_{\text{ols}} = (\wh\gamma_{\text{ols},1},\dots, \wh \gamma_{\text{ols},q})^\top \in\mR^q$. For convenience, define $\bm \theta = (\bm \beta^\top, \bm \gamma^\top)^\top \in \mR^{p+q}$, $\wh {\bm \theta}_{\text{ols}} = (\wh{\bm \beta}^\top_{\text{ols}}, \wh{\bm \gamma}^\top_{\text{ols}})^\top \in \mR^{p+q}$. Then we should have $\wh{\bm \theta}_{\text{ols}} = (\sum_{i=1}^N \bm U_i\bm  U_i^\top)^{-1}\times (\sum_{i=1}^N \bm U_iY_i) $, where $\bm U_i = (\bm Z_i^\top, \bm X_i^\top)^\top\in\mR^{p+q}$. If all the variables are fully observed, the OLS estimator will be consistent and asymptotically normal under mild regularity conditions \citep{rao1973linear,shao2003mathematical}.

Unfortunately, in this interested application, the binary features $\bm Z_i$ are largely missing because of the high collection cost. To solve the problem, we assume an auxiliary variable $\bm W_i = (W_{i1}, \dots, W_{ir})^\top \in\mR^r$ could be collected for the $i$th observation. We typically expect $\bm W_i$ to be highly related to $\bm Z_i$ and to be much easier to collect. To model the relationship between $\bm Z_i$ and $\bm W_i$, a logistic regression model is correctly specified as follows
\[
P\Big(Z_{ij} = 1\big|\bm W_i\Big) = \frac{\exp(\bm W_i^\top \bm \alpha_j)}{1+\exp(\bm W_i^\top \bm \alpha_j)} = p\big(\bm W_i^\top\bm \alpha_j\big),
\]
where $\bm \alpha_j= (\alpha_{j1},\dots,\alpha_{jr})^\top \in \mR^r(1\le j \le p)$ is the coefficient vector and $p(x) = \exp(x)/\{1+\exp(x)\}$ is the sigmoid function. For convenience, we refer to this model as an \textit{imputation model}. We also assume that $Z_{ij_1}$ and $Z_{ij_2}$ are conditionally independent on $\bm W_i$ for $j_1\neq j_2$. If a consistent estimator for $\bm \alpha_j$ with the desired statistical accuracy can be obtained, the binary vector $\bm Z_i$ could be predicted with reasonable accuracy. More specifically, we first extract a subsample out of the whole dataset. Without loss of generality, we assume the first $n$ observations are selected. We typically expect $n$ to be much smaller than $N$ in the sense $n / N \to 0$ as $N \to \infty$. We assume that for each $1\le i\le n$ the accurate value of $\bm Z_i$ could be collected. Obviously, this leads to a considerable data collection cost. However, as $n\ll N$, the cost for collecting $\bm Z_i$ with $1\le i\le n$ is expected to be practically acceptable. For convenience, we refer to $\mS_0 = \{(\bm Z_i, \bm W_i, \bm X_i, Y_i ): 1\le i\le n\}$ as the pilot sample.

With the help of the pilot sample $\mS_0$, the unknown regression coefficient $\bm \alpha_j$ can be consistently estimated by maximizing the log-likelihood function
\[
\mL_{j,z}(\bm \alpha_j;\mS_0) = \sum_{i=1}^n\bigg[ Z_{ij} \bm W_i^\top \bm \alpha_j - \log \Big\{1+\exp\Big(\bm W_i^\top \bm \alpha_j\Big)\Big\}\bigg].
\]
Then the maximum likelihood estimator for $\bm \alpha_j$ is given as $\wh{\bm \alpha}_j =\argmax_{\bm \alpha_j}\mL_{j,z}(\bm \alpha_j)$. Under appropriate regularity conditions, the maximum likelihood estimator $\wh{\bm \alpha}_j$ is consistent and asymptotically normal \citep{shao2003mathematical,lehmann2006theory}. We impute the binary covariate $\bm Z_i$ by its predicted probability $\wh {\bm Z}_i$, where the $j$th component of $\wh {\bm Z_i}$ is given as $\wh Z_{ij} = p(\bm W_i^\top \wh {\bm \alpha}_j)$. Here we do not impute $\wh Z_{ij}$ by its binary prediction results (e.g. $\wh{Z}_{ij}=I\{p(\bW_i^\top \wh \balpha_j)>0.5\}$) due to the challenge of threshold value selection. As pointed out by \cite{qiao2009adaptive}, when $\bZ_i$ is highly imbalanced, the threshold value 0.5 is unlikely to be the optimal choice. In the meanwhile, what is the optimal choice about the threshold value is not immediately clear.

After imputation, we obtain an imputed whole sample as $\mS_{\text{imp}} = \mS_0 \cup \mS_1$, where $\mS_1 = \{(\wh {\bm Z}_i, \bm  W_i, \bm X_i, Y_i): n+1\le i\le N\}$. Next a loss function can be constructed based on the imputed whole sample as
\beq
\mL_{N,y}(\bm \beta,\bm \gamma;\mS_{\text{imp}}) = \sum_{i=1}^n \Big(Y_i - \bm Z_i^\top \bm \beta - \bm X_i^\top \bm \gamma\Big)^2 + \sum_{i=n+1}^{N} \Big(Y_i - \wh {\bm Z}_i^\top \bm \beta - \bm X_i^\top \bm \gamma\Big)^2.\label{OLS with imputed variables}
\eeq
Then, an imputed whole sample based estimator can be easily obtained as $(\wh{\bm \beta}_{\text{imp}},\wh{\bm \gamma}_{\text{imp}}) = \argmin_{\bm \beta,\bm \gamma} \mL_{N,y}(\bm \beta,\bm \gamma; \mS_{\text{imp}})$. For convenience, we refer to the estimator $\wh {\bm \theta}_{\text{imp}} = (\wh {\bm \beta}_{\text{imp}}^\top, \wh{\bm \gamma}_{\text{imp}}^\top)^\top \in \mR^{p+q}$ with $\wh {\bm \beta}_{\text{imp}} = (\wh \beta_{\text{imp},1},\dots,\wh \beta_{\text{imp},p})^\top\in\mR^p$ and $\wh {\bm \gamma}_{\text{imp}} = (\wh\gamma_{\text{imp},1},\dots,\wh \gamma_{\text{imp},q})^\top\in\mR^q$ as the \textit{imputed estimator}. Further, we define $\wh {\bm U}_i = (\wh {\bm Z}_i^\top, \bm X_i^\top )^\top\in \mR^{p+q}$. It can be easily verified that $\wh {\bm \theta}_{\text{imp}} = (\sum_{i=1}^n\bm U_i\bm U_i^\top +\sum_{i=n+1}^N\wh {\bm U}_i\wh {\bm U}_i^\top)^{-1}(\sum_{i=1}^n \bm U_iY_i + \sum_{i=n+1}^N \wh {\bm U}_i Y_i)$. The asymptotic properties of the estimator are to be studied subsequently. We investigate the theoretical properties in the regular case first and then discuss the two special cases, which are the highly imbalanced and highly predictable cases.

It is remarkable that two models are included here. The first one is the true model \eqref{linear regression}, where $\bZ_i$'s are all binary. The second one is the working model \eqref{OLS with imputed variables}, where the binary $\bZ_i$'s are partially imputed by the estimated response probabilities. The imputed response probability is continuous. The regression coefficient $\bbeta$ should be interpreted with respect to the true model, where the $\bZ_i$'s are all binary. The working model \eqref{OLS with imputed variables} is developed solely for the purpose of parameter estimation rather than interpretation.

\subsection{The Asymptotic Theory for Regular Case}

To study the asymptotic properties of the imputed estimator $\wh{\bm \theta}_{\text{imp}}$, we start with the \textit{regular case} with fixed $\bm \alpha_j$'s, which means the binary covariates are relatively balanced and the prediction accuracy of the covariates is not extremely high. Accordingly, we should never have $p(\bm W_i^\top\bm \alpha_j) = Z_{ij}$ or even approximately in any sense. In other words, even if the true parameter $\bm \alpha_j$ is given, we can never predict $\bm Z_i$ consistently. Therefore, the consistency of $\wh {\bm \theta}_{\text{imp}}$ becomes skeptical \citep{white2010bias}. Let $\mA = (\bm \alpha_1^\top,\dots,\bm \alpha_p^\top)^\top\in \mR^{pr}$ and $\wh \mA = (\wh{\bm \alpha}_1^\top,\dots,\wh {\bm \alpha}_p^\top)^\top \in \mR^{pr}$ be the maximum likelihood estimator of $\mA$. Let $\bm p_i = (p(\bm W_i^\top\bm \alpha_1),\dots, p(\bm W_i^\top\bm \alpha_p))^\top \in \mR^p$ and $\wt {\bm U}_i = (\bm p_i^\top, \bm X_i^\top)^\top \in \mR^{p+q}$. Define $\mathbb D_{ip} = \diag[p(\bm W_i^\top\bm \alpha_1)\{1-p(\bm W_i^\top\bm \alpha_1)\},\dots,p(\bm W_i^\top\bm \alpha_p)\{1-p(\bm W_i^\top\bm \alpha_p)\}] \in \mR^{p\times p}$ and $\mW = (\mE(\mathbb D_{ip}),\mathbf 0_{p\times q};\mathbf 0_{q\times p},\mathbf 0_{q\times q}) \in \mR^{(p+q)\times (p+q)}$. To investigate the asymptotic behavior of $\wh{\bm \theta}_{\text{imp}}$, the following assumptions are required.

\begin{itemize}
	\item [(C1)] (Sample divergence rate) As $N\to\infty$, we assume that $n \to\infty$ and $n/N\to0$.
	\item [(C2)] (Nonsingular matrices) Assume that the matrices $\Sigma_u = \mE(\bm U_i\bm U_i^\top)$, $\Sigma_u - \mW$, and $\mE[\mathbb D_{ip} \otimes (\bm W_i\bm W_i^\top)]$ are finite and positive definite.
\end{itemize}

%

\noindent For condition (C1), we should expect the pilot sample size $n$ to be much smaller than the total sample size $N$ in the sense that $n/N\to 0$ as $N\to\infty$. This indicates that the data we have labeled is only a small portion of the entire dataset. Condition (C2) is a classical assumption to ensure the nonsingularity of both the Fisher information matrix in the imputation model and the covariance matrix in the substantive model. Under these conditions, we have the following theorem.

\begin{theorem}
	Assume conditions (C1) and (C2), we have $\wh{\bm \theta}_{\text{imp}} - \bm \theta = (\Sigma_u - \mW)^{-1}(\bm \zeta_1 + \bm \zeta_2)\{1+o_p(1)\}$. 
	Moreover, $\bm \zeta_1$ and $\bm \zeta_2$ are two independent random quantities such that $\bm \zeta_1 = -\Sigma_{uw}(\wh \mA - \mA)$ and $\sqrt N\bm \zeta_2 \stackrel{d}{\to} N(0,\Omega)$, where $\Sigma_{uw} = \mE\{(\bm \beta^\top \mathbb D_{ip})\otimes( \bm U_i \bm W_i^\top) \}$, and $\Omega = \mE[\{\bm \beta^\top\mathbb D_{ip}\bm \beta + \sigma^2 \} \wt {\bm U}_i\wt {\bm U}_i^\top]$.
	\label{Theorem regular}
\end{theorem}

\noindent The detailed proof of Theorem \ref{Theorem regular} is provided in Appendix B in the supplementary material. By Theorem \ref{Theorem regular}, we know that the imputed estimator $\wh {\bm \theta}_{\text{imp}}$ is consistent. Moreover, the estimation error $\wh{\bm \theta}_{\text{imp}} - \bm \theta$ can be decomposed into two parts. The first part $\bm \zeta_1$ is a linear transformation of $\wh {\bm \alpha}_j - \bm \alpha_j$. Therefore, its asymptotic behavior is fully determined by the estimation error due to the imputation model. Consequently, we know that $\bm \zeta_1$ is an $O_p(n^{-1/2})$ term and is asymptotically normal. The second part $\bm \zeta_2$ involves errors due to both imputation and substantive models. Under the assumption $n/N\to0$, the second term $\bm \zeta_2$ is an $O_p(N^{-1/2})$ term and therefore is asymptotically negligible when compared with the first term $\bm \zeta_1$. Combining these two facts, we know that $\wh {\bm \theta}_{\text{imp}}$ is $\sqrt n$-consistent for $\bm \theta$.
This is a convergence rate the same as that of the OLS estimator obtained by using the pilot sample only, that is $\wh {\bm \theta}_{\text{pilot}} = (\wh {\bm \beta}_{\text{pilot}}^\top, \wh {\bm \gamma}_{\text{pilot}}^\top)^\top = (\sum_{i=1}^n \bm U_i\bm U_i^\top)^{-1}(\sum_{i=1}^n\bm U_i Y_i)$, where $\wh {\bm \beta}_{\text{pilot}} = (\wh\beta_{\text{pilot},1}, \dots,\wh \beta_{\text{pilot},p})^\top \in\mR^p$ and $\wh{\bm \gamma}_{\text{pilot}} = (\wh \gamma_{\text{pilot},1}, \dots,\wh \gamma_{\text{pilot},q})^\top \in\mR^q$. For convenience, we refer to this estimator as the \textit{pilot estimator}. Under appropriate regularity conditions, we can verify that $\sqrt{n}(\wh{\bm \theta}_{\text{pilot}} - \bm \theta)\stackrel{d}{\to}N(0,\sigma^2\Sigma_u^{-1})$.

Note that the two estimators $\wh{\bm \theta}_{\text{imp}}$ and $\wh{\bm \theta}_{\text{pilot}}$ have the same convergence rate, but very different asymptotic covariance matrices. Then a natural question is: Which estimator is better? We are particularly interested in the situation when $\wh {\bm \theta}_{\text{imp}}$ outperforms $\wh {\bm \theta}_{\text{pilot}}$, as this is the case, the imputation efforts are not wasted. By Theorem \ref{Theorem regular}, we know that the asymptotic efficiency of $\wh {\bm \theta}_{\text{imp}}$ is mainly determined by that of $\wh \mA$. By contrast, that of $\wh {\bm \theta}_{\text{pilot}}$ is significantly affected by the variance of the random noise $\sigma^2$. It is remarkable that this quantity $\sigma^2$ is not involved in the leading asymptotic covariance of $\wh{\bm \theta}_{\text{imp}}$. In fact, the variance $\sigma^2$ affects the asymptotic variance of $\wh {\bm \theta}_{\text{imp}}$ through $\bm \zeta_2$. See Step 5 of Appendix B for the detailed expression for $\bm \zeta_2$. However, $\bm \zeta_2 = o_p(\bm \zeta_1)$ under  condition (C1) is thus ignorable. In this way, $\sigma^2$ does not affect the leading covariance of $\wh{\bm \theta}_{\text{imp}}$. This discussion suggests that $\wh {\bm \theta}_{\text{imp}}$ should be a better choice than $\wh{\bm \theta}_{\text{pilot}}$ if $\sigma^2$ is relatively large. Otherwise, it might not be a good choice for a larger computational cost. In the extreme case with $\sigma^2 = 0$, we should have $\wh{\bm \theta}_{\text{pilot}} = \bm \theta$. Then no imputation is needed. 
\subsection{Special Case I: The Highly Imbalanced Case}

Theorem \ref{Theorem regular} studies the asymptotic behavior of $\wh{\bm \theta}_{\text{imp}}$ for regular cases, where the response probability needs to be relatively balanced. In other words, we should have both $\min_j \mE\{p(\bm W_i^\top\bm \alpha_j)\}$ and $1-\max_j \mE\{p(\bm W_i^\top\bm \alpha_j)\}$ well bounded above zero. However, in real practice, we often encounter the situation where positive cases in $\bm Z$ are rare. 
For example, for the pharmacokinetic analysis of tacrolimus, the percentage of immunoglobulin treatment accounts for approximately only 7.9\% of the concentration records \citep{chen2017population}. Theoretically, this suggests that it might be more appropriate to assume that $P(Z_{ij} = 1) \to 0$ as the sample size $N\to\infty$. We next study the asymptotic behavior of $\wh{\bm \theta}_{\text{imp}}$ in this case in this subsection. To this end, we follow the theoretical framework of \cite{wang2020logistic} and \cite{wang2021nonuniform} and write $\bm W_i = (1,\wt {\bm W}_i^\top)^\top\in \mR^r$ and $\bm \alpha_j = (\alpha_{Nj}, \bm \alpha_j^{*\top})^\top\in\mR^r$ with $\wt {\bm W}_i \in \mR^{r-1}$ to be the subvector of $\bm W_i$, $\alpha_{Nj} \in \mR$ and $\bm \alpha_j^* \in \mR^{r-1}$. Define $\bm \pi_i = (\exp(\wt {\bm W}_i^\top \bm \alpha_1^*),\dots,\exp(\wt {\bm W}_i^\top \bm \alpha_p^*))^\top \in \mR^{p}$. Let $V_u = (V_z, 0;0, \Sigma_x)$ with $V_z = \mE\{\diag(\bm \pi_i)\}\in\mR^{p\times p}$ and
$\Sigma_x = \mE(\bm X_i\bm X_i^\top) \in \mR^{q\times q}$. Further let $\wt V_u = (\wt V_z, V_{zx}; V_{xz}, \Sigma_x)$, $\wt V_z = \mE(\bm \pi_i\bm \pi_i^\top) \in\mR^{p\times p}$, 
$V_{zx} = V_{xz}^\top = \mE(\bm \pi_i \bm X_i^\top)\in\mR^{p\times q}$. Define $r_{Nj} = \exp(\alpha_{Nj})$ be the positive proportion for the binary variable $Z_{ij}, j = 1,\dots, p$. Let $r_{N,\max} = \max_j r_{Nj} =  \max_j \exp(\alpha_{Nj})$ and $r_{N,\min} = \min_j r_{Nj} = \min_j \exp(\alpha_{Nj})$ be the maximum and minimum positive proportion for the binary vector $\bm Z_i$, respectively. To investigate the asymptotic behavior of the estimators, the following conditions are needed.

\begin{itemize}
	\item [(C3)] (Imbalance effect) As $N\to\infty$, we assume that  $r_{Nj} \to 0$, $n r_{Nj} \to\infty$, and $r_{Nj} / r_{N,\max} \to c_j$ for some constants $c_j \in [0,1]$ for $1\le j\le p$.
	\item [(C4)] (Existence of moment generating function) Assume that $\mE\exp(\tau \|\wt {\bm W}_i\|) <\infty$ for all $\tau >0$.
	\item [(C5)] (Nonsingular matrices) Assume that $V_u$, $\wt V_u$ and $\mE\{\diag(\bm \pi_i) \otimes (\bm W_i\bm W_i^\top) \}$ are finite and positive definite.
\end{itemize}

\noindent Condition (C3) formulates the case where all the binary covariates have rare positive cases. By assuming $r _{Nj}\to 0$, we know that $P(Z_{ij} = 1) = r_{Nj}\mE[\exp(\wt {\bm W}_i^\top \bm \alpha_j^*) / \{1+\exp(\alpha_{Nj} + \wt {\bm W}_i^\top \bm \alpha_j^*)\}] \to 0$ as $N\to\infty$. Therefore, the imbalanced phenomenon can be theoretically described. Moreover, by assuming $nr_{Nj}\to \infty$, we know that $\mE(\sum_{i=1}^n Z_{ij}) = nr_{Nj}\mE[\exp(\wt {\bm W}_i^\top \bm \alpha_j^*) / \{1+\exp(\alpha_{Nj} + \wt {\bm W}_i^\top \bm \alpha_j^*)\}] \to\infty$. This suggests that the number of positive cases should diverge to infinity, even if its sample percentage is low. This allows us to develop rigorous asymptotic theory. Condition (C4) requires the existence of the moment generating function for $\wt \bW_i$. This condition ensures the existence of a dominated function when applying the dominated convergence theorem in the proof. Condition (C5) is similar to condition (C2), which ensures the positive definiteness of the asymptotic covariance matrices. It is remarkable that condition (C3) is about the intercept parameter, which is the scalar $\alpha_{Nj} = \log (r_{Nj}) \to -\infty$ as $N\to\infty$. In contrast, the condition (C5) is about $V_u$, $\wt V_u$ and $V_\pi = \mE\{\diag (\bm \pi_i)\otimes (\bW_i \bW_i^\top)\}$, which is basically a technical condition for regular feature vectors $\bX_i$ and $\wt \bW_i$. Note that $\bX_i$ and $\wt \bW_i$ are regular feature vectors with finite dimensions. As a result, the positive definiteness of the covariance matrices in (C5) is naturally satisfied.
Under the aforementioned assumptions, we further define $\mathbb D = \diag\{r_{N1}, \dots, r_{Np}\} \in \mR^{p\times p}$ and $\wt {\mathbb D} = \diag\{\mathbb D, I_q \} \in \mR^{(p+q)\times(p+q)}$, where $I_q$ is a $q\times q$ identity matrix. We then obtain the following theorem about the theoretical properties of both $\wh{\bm \theta}_{\text{pilot}}$ and $\wh{\bm \theta}_{\text{imp}}$.

\begin{theorem}
	(1) (\textbf{Asymptotic distribution of the pilot estimator.}) Assume conditions (C1) and (C3)-(C5)
	, we then have $\sqrt n\wbD^{1/2}(\wh{\bm \theta}_{\text{pilot}} - \bm \theta) = (\sqrt n\mathbb D^{1/2} (\wh{\bm \beta}_{\text{pilot}} - \bm \beta)^\top, \sqrt n(\wh{\bm \gamma}_{\text{pilot}} - \bm \gamma)^\top)^\top \stackrel{d}{\to} N(0,\sigma^2 V_u^{-1})$;
	
	(2) (\textbf{Asymptotic distribution of the imputed estimator.}) Further assume that $Nr_{N,\min}/n\to\infty$ as $N\to\infty$, 
	we have $\sqrt{nr_{N,\max}^{-1}} \wbD(\wh{\bm \theta}_{\text{imp}} - \bm \theta) = (\sqrt{nr_{N,\max}^{-1}}\mathbb D(\wh{\bm \beta}_{\text{imp}} - \bm \beta)^\top, \sqrt{n r_{N,\max}^{-1}}(\wh{\bm \gamma}_{\text{imp}} - \bm \gamma)^\top)^\top \stackrel{d}{\to} N(0,V_{\text{imp}})$, where $V_{\text{imp}} = \wt V_u^{-1}(\sum_{j=1}^{p} c_j\beta_j^2 V_{\pi w_j}V_jV_{\pi w_j}^\top) \wt V_u^{-1}$ with the matrices $V_{\pi w_j} = \mE\{\exp(\wt {\bm W}_i^\top\bm \alpha_j^*)\bm \pi_i\bm W_i^\top\}$ and $V_j = \mE\{\exp(\wt {\bm W}_i^\top\bm \alpha_j^*) \bm W_i\bm W_i^\top\}^{-1}$.

	\label{Theorem imbalanced}
\end{theorem}
\noindent The detailed proof of Theorem \ref{Theorem imbalanced} is provided in Appendix C in the supplementary material. By Theorem \ref{Theorem imbalanced}, we know that the data imbalance does play a critical role in the convergence rate of the corresponding estimator. We discuss this for the estimation of $\bm \beta$ and $\bm \gamma$ separately.

\noindent (1) \textbf{Estimation of $\bm \beta$}. First, for the pilot estimator, by Theorem \ref{Theorem imbalanced}, we know that $\wh{\bm \beta}_{\text{pilot},j} - \bm \beta_j = O_p(n^{-1/2}r_{Nj}^{-1/2})$. Therefore, we know that in the highly imbalanced case, the convergence rate of $\wh{\bm \beta}_{\text{pilot}}$ will be slower than the classical parametric rate $\sqrt n$. Second, for the imputed estimator, we know that $\wh\beta_{\text{imp},j} - \beta_j = O_p(n^{-1/2}r_{N,\max}^{1/2}r_{Nj}^{-1})$. The convergence rate can be studied according to two cases. \textbf{Case 1}: If the $Z_{ij}$s are typically balanced with $c_j>0$, then we have $\wh\beta_{\text{imp},j} - \beta_j = O_p(n^{-1/2}r_{Nj}^{-1/2})$, which is the same convergence rate as that of $\wh\beta_{\text{pilot},j}$. \textbf{Case 2}: If the $Z_{ij}$s are extremely imbalanced with $c_j = 0$, then we have $\wh\beta_{\text{imp},j} - \beta = O_p(n^{-1/2}r_{N,\max}^{1/2}r_{Nj}^{-1})$, which is a convergence rate slower than that of $\wh\beta_{\text{pilot},j}$.

To gain some intuitive understanding about Case 2, we consider here a highly simplified example as $Y_i = Z_{i1}\beta_1 + Z_{i2} \beta_2 + \ve_i$, where $P(Z_{i1}= 1) = 1/2$ and $P(Z_{i2} = 1) \approx r_{N2}\mE\{\exp(\wt \bW_i^\top\balpha_2^* )\} \to 0$, where $r_{N2}= \exp(\alpha_{N2}) \to 0$ as $N\to\infty$. For simplicity, we write $r_N = r_{N2}$ in this simplified example. Next we compare the performance of pilot and imputed estimators. We have $Z_{i2}$ actually observed for the pilot sample. In this case, the information contained in $Z_{i2}$, as measured by its variance, is given by $\var(Z_{i2}) = \mE\{p_{i2}(1-p_{i2})\} = O(r_N)$ since $p_{i2} = P(Z_{i2} =1 |\bW_i) \stackrel{p}{\to} 0$ as $N\to\infty$. However, the amount of information contained in $\wh Z_{i2}$ is given by $\var(\wh Z_{i2}) \le \mE(\wh Z_{i2}^2) \approx \mE(p_{i2}^2) = O(r_N^2)$, which is a smaller order term as compared with $\var(Z_{i2})$. Intuitively, this suggests that a significant amount of variability about $Z_{i2}$ is lost with $\wh Z_{i2}$, if the interested case is a highly imbalanced one. That intuitively explains the slower convergence rate of $\wh\beta_{2,\text{imp}}$.

\noindent (2) \textbf{Estimation of $\bm \gamma$.} We next study the asymptotic behavior of $\wh{\bm \gamma}_{\text{imp}}$ and $\wh{\bm \gamma}_{\text{pilot}}$.  From Theorem \ref{Theorem imbalanced}, we know that the convergence rate of $\wh{\bm \gamma}_{\text{pilot}}$ is $O_p(n^{-1/2})$, which is the same as the regular case. By contrast, we have $ \wh {\bm \gamma}_{\text{imp}} - \bm \gamma = O_p(n^{-1/2}r_{N,\max}^{1/2})$, which is a convergence rate faster than that of $\wh{\bm \gamma}_{\text{pilot}}$ since $r_{N,\max} = \max_jr_{Nj}\to 0$ as $N\to\infty$. Consequently, we should expect that $\wh{\bm \gamma}_{\text{imp}}$ has a smaller variance than $\wh{\bm \gamma}_{\text{pilot}}$.

To gain some intuitive understanding of  the faster convergence rate of $\wh\bgamma_{\text{imp}}$, we consider here a highly simplified model as $Y_i = Z_i\beta + X_i\gamma + \ve_i$, where $Z_i$ is a highly imbalanced binary variable with $P(Z_i = 1) \approx r_N\mE\{\exp(\wt \bW_i^\top \balpha^*)\}  \to 0$ with $r_N = \exp(\alpha_{N})$ and $X_i$ {\blue is} a standard normal random variable independent of $Z_i$. With imputed feature $\wh Z_i$, we can then rewrite the regression model as $Y_i = \wh Z_i \beta + X_i\gamma + (Z_i - \wh Z_i)\beta + \ve_i$. One can verify that the side effect for $\wh\bgamma_{\text{imp}}$ due to imputation error is mainly caused by the term $\{N^{-1}\sum_{i=1}^N X_ip_i(1-p_i)\bW_i\}^\top (\balpha - \wh\balpha) =  r_N\mE\{\exp(\wt\bW_i^\top \balpha^*) X_i \bW_i\}^\top (\balpha - \wh\balpha)\{1+o_p(1)\}$. For the binary variable $Z_i$, the more imbalanced it is, the larger estimation error it suffers from $\balpha - \wh\balpha = O_p(1/\sqrt{nr_N})$. Nevertheless, this side effect can be significantly discounted by a factor of $r_N$ due to the fact that $\var(Z_i) = \mE\{p_i(1-p_i)\} = O(r_N)$. To fix the idea, consider, for example, the most extreme situation with $P(Z_i = 0) = 1$. We then have $\var(Z_i) = \mE\{p_i(1-p_i)\} =0$ and $r_N = 0$. The consequence is that $N^{-1}\sum_{i=1}^N X_ip_i(1-p_i)\bW_i = 0$. In this case, no imputation error can be transfer from $\wh\balpha$ to $\wh\gamma_{\text{imp}}$. Therefore, the overall side effect due to imputation also disappears. This explains why overall speaking less imbalanced terms are more influential in affecting the convergence rate of $\wh\bgamma_{\text{imp}}$.

\subsection{Special Case II: The Highly Predictable Case}

Other than the regular case studied in the previous subsection, we often encounter practical cases with extremely high prediction accuracy, which means either $p(\bm W_i^\top\bm \alpha_j) \to 0$ or $p(\bm W_i^\top\bm \alpha_j )\to 1$. In this case, the difference between the estimated probability $p(\bm W_i^\top\wh {\bm \alpha}_j)$ and the actual binary feature $\bm Z_i$ is asymptotically ignorable. Mathematically, this can be represented by imposing a technical condition that $p(\bm W_i^\top \bm \alpha_j)\{1-p(\bm W_i^\top\bm  \alpha_j)\}\to 0$ in the sense that $N\to\infty$. A sufficient but necessary condition for this condition is that $\|\bm \alpha_j\| \to\infty$ at an appropriate speed as $N\to\infty$. 

We are inspired to consider this interesting case mainly because, 
for many modern machine learning related classification problems, the out-of-sample prediction accuracy can be extremely high. For illustration, consider for example the CIFAR-10 classification problem with 10 different classes \citep{krizhevsky2009learning}. An excellent prediction accuracy of 99.50\% has been achieved by \cite{dosovitskiy2020image}. As another example, consider the Oxford Flower dataset with a total of 102 classes \citep{nilsback2008automated}. An outstanding forecasting accuracy of 99.76\% has been achieved by \cite{hassani2021escaping}. For these applications, we should always have $\max_j\mE[p(\bm W_i^\top\bm \alpha_j)\{1-p(\bm W_i^\top\bm \alpha_j)\}]$ extremely close to 0. This makes the asymptotic behavior of the related statistical estimators, such as the maximum likelihood estimator for the logistic regression, somewhat different from the classical cases. 

Therefore, we are motivated to study the asymptotic behavior of the proposed imputed estimator under this highly predictable case. Define $\omega_n =\max_j\mE[p(\bm W_i^\top\bm \alpha_j)\{1-p(\bm W_i^\top\bm \alpha_j)\}]$ to be the maximum prediction variation of $\bm Z_i$. Then a smaller value of $\omega_n$ implies a higher prediction accuracy. One can verify that if $\|\bm \alpha_j\| \to \infty$ for all $1\le j\le p$, we have $\omega_n \to 0$ as $N\to\infty$. Specifically, the following assumptions are required.

\begin{itemize}
	\item [(C6)] (Predictability of binary covariates) As $N\to\infty$, we assume that $\omega_n\to0$, $n/(N\omega_n) \to \infty$, and $n\omega_{n} \to\infty$.
	\item [(C7)] (Local dominance function) Define $\mathbb B(\bm \alpha,r) = \{\wt {\bm \alpha}: \|\wt {\bm \alpha} - \bm \alpha\|\le r\}$ as a compact local ball with $\bm \alpha$ as the center and $r>0$ as the radius. Then we assume that there exists a sufficiently small but fixed $r$ such that $\max_j \mE[\sup_{\bm \alpha\in \mathbb B(\bm \alpha_j,r)} p(\bm W_i^\top\bm \alpha)\{1-p(\bm W_i^\top\bm \alpha_j)\}\{\|\bm W_i\|^6 + \|\bm X_i\|^2\}]\le C_{\max}\omega_n$, where $C_{\max} >0$ is some fixed constant. 
	\item[(C8)] (Minimum eigenvalue) Assume there exists a fixed constant $0<C_{\min}<\infty $ such that $\lambda_{\min}(\mE[p(\bm W_i^\top\bm \alpha_j)\{1-p(\bm W_i^\top\bm \alpha_j)\} \bm W_i \bm W_i^\top]) \ge C_{\min }\omega_n$ for all $1\le j\le p$, where $\lambda_{\min}(A)$ represents the minimum eigenvalue of an arbitrary symmetric matrix $A$.
\end{itemize}
Recall that the pilot sample size $n$ is assumed to be much smaller than the total sample size $N$ in the sense that $n/N\to 0$ as $N\to\infty$ in condition (C1). To ensure $n / (N\omega_n ) \to\infty$ as $N\to\infty$ in (C6), we must have $\omega_n\to 0$ at a sufficiently fast speed. This condition further implies that for an arbitrary case with a feature vector $\bm W_i$, we should have $p(\bm W_i^\top\bm \alpha_j)$ very close to either 1 or 0. Otherwise, we cannot have $p(\bm W_i^\top\bm \alpha_j)\{1-p(\bm W_i^\top\bm \alpha_j)\}$ close to 0.
However, the assumption $n\omega_n \to \infty$ as $N\to\infty$ in (C6) constrains the convergence rate of $\omega_n$ toward 0 to not be too fast either. Otherwise the assumption $n\omega_n \to\infty$ as $N\to\infty$ would be violated. This assumption implies that the number of both positive and negative cases in the pilot sample should diverge to infinity as $n\to\infty$. To see this, note that the expected number of positive cases in the pilot sample is given by $n\mE\{p(\bm W_i^\top\bm \alpha_j)\}\ge n\mE[p(\bm W_i^\top\bm \alpha_j)\{1-p(\bm W_i^\top\bm \alpha_j)\}] = n\omega_n$. Similarly, the expected number of negative cases in the pilot sample is given by $n\mE\{1-p(\bm W_i^\top\bm \alpha_j)\} \ge n\mE[p(\bm W_i^\top\bm \alpha_j)\{1-p(\bm W_i^\top\bm \alpha_j)\}] \ge n\omega_n$. Then by the condition that $n\omega_n \to \infty$ as $N\to\infty$ in (C6), we know that both the number of positive and negative cases in the pilot sample should diverge to infinity as $N\to\infty$. This is also a reasonable assumption. Otherwise, no statistically consistent estimator can be computed based on the pilot sample. Combining the conditions $n/(N\omega_n) \to\infty$ and $n\omega_n \to\infty$ as $N\to\infty$, we implicitly assume that $n^2/ N\to\infty$. In this way, this assumption further requires that the pilot sample size cannot be too small either.

Condition (C7) is basically a moment type condition. It constrains the convergence rate of some $p(\bm W_i^\top\bm \alpha_j)\{1-p(\bm W_i^\top\bm \alpha_j)\}$ related moments to be no slower than $\omega_n$. The last condition (C8) constrains the information contained in the information matrix $\mE[p(\bm W_i^\top\bm \alpha_j)\{1-p(\bm W_i^\top\bm \alpha_j)\}\bm W_i\bm W_i^\top]$ to not be too small. With the help of these assumptions, we then have the following theorem about $\wh{\bm \theta}_{\text{imp}}$.

\begin{theorem}
	Under conditions (C1)-(C2) and (C6)-(C8), we have $\sqrt N(\wh {\bm \theta}_{\text{imp}} - \bm \theta) \stackrel{d}{\to} N(0,\sigma^2\Sigma_u^{-1})$\label{Theorem high accuracy}.
\end{theorem}
\noindent The detailed proof of Theorem \ref{Theorem high accuracy} is provided in Appendix D in the supplementary material. By Theorem \ref{Theorem high accuracy}, we know that, if we can impute the missing variable $\bm Z_i$ with sufficiently high accuracy, the asymptotic behavior of $\wh{\bm \theta}_{\text{imp}}$ could be the same as that of the oracle OLS estimator $\wh{\bm \theta}_{\text{ols}}$, which is the estimator obtained with all $\bm Z_i$ values observed. It is then of great interest to further explore the intrinsic relationship between Theorems \ref{Theorem regular} and \ref{Theorem high accuracy}. By Theorem \ref{Theorem regular}, we know that $\wh{\bm \theta}_{\text{imp}} - \bm \theta$ can be decomposed as $\wh{\bm\theta}_{\text{imp}} - \bm \theta = (\Sigma_u - \mW)^{-1}(\bm \zeta_1 + \bm \zeta_2)\{1+o_p(1)\}$ with $\sqrt N \bm \zeta_2 \stackrel{d}{\to} N(0,\Omega)$. If the same technical condition as in Theorem \ref{Theorem high accuracy} can be assumed, we then have (1) $\Sigma_u - \mW = \Sigma_u\{1+o(1)\}$, (2) $\bm \zeta_1 = o_p(\bm \zeta_2)$, and (3) $\Omega = \sigma^2\Sigma_u\{1+o(1)\}$. The verification details are given in Appendix D. It follows then that $\sqrt N (\wh{\bm \theta}_{\text{imp}} - \bm \theta) = \sqrt N \Sigma_u^{-1}\bm \zeta_2 + o_p(1) \stackrel{d}{\to} N(0,\sigma^2 \Sigma_u^{-1})$, which is the same conclusion as given in Theorem \ref{Theorem high accuracy}.

\subsection{Unified Covariance Matrix Estimation}

We next consider how to estimate the asymptotic covariance of $\wh{\bm \theta}_{\text{imp}}$. By Theorem \ref{Theorem regular}, we know that the asymptotic covariance matrix of the imputed estimator is considerably more complicated than that of a standard estimator without imputation. By treating the imputed values as if they were the truth, the estimated asymptotic covariance will be seriously biased, since we are trying to estimate the asymptotic covariance of the oracle OLS estimator $\wh\btheta_{\text{ols}}$. As a result, the standard covariance estimator assuming imputed values are true fails. In the meanwhile, the previous analysis suggests that different cases of belonging lead to different analytical formulas according to the case that the intended application belongs to. 
For convenience, it is of great interest to have a unified asymptotic covariance estimator, which works well for all the cases. Specifically, we start with the asymptotic covariance estimation for the regular case. By Theorem \ref{Theorem regular}, we know that the asymptotic covariance of $\wh{\bm \theta}_{\text{imp}}$ is mainly determined by the following quantities. They are $\Sigma_u - \mW$, $\Sigma_{uw} = \mE\{(\bm \beta^\top \mathbb D_{ip})\otimes( \bm U_i \bm W_i^\top) \}$, $\sigma^2$, $\Omega$ and the asymptotic covariance matrix of $\wh \mA - \mA$, which is given by $\mI(\mA) = 
\mE\{\mathbb D_{ip} \otimes (W_iW_i^\top)\}^{-1}$. Accordingly, they can be estimated based on the pilot sample as
\begin{gather*}
	\wt \Sigma_{u} = n^{-1}\sum_{i=1}^n \wh {\bm U}_i\wh {\bm U}_i^\top,
	\wh\Sigma_{uw} =n^{-1}\sum_{i=1}^n \Big(\wh {\bm \beta}_{\text{pilot}}^\top \wh {\mathbb D}_{ip}\Big) \otimes \Big( \bm U_i \bm W_i^\top\Big),\\
	\wh \sigma^2_{\text{pilot}} = (n - p - q)^{-1}\sum_{i=1}^n\Big(Y_i - \bm U_i^\top\wh{\bm \theta}_{\text{pilot}}\Big)^2,
	\wh \Omega = n^{-1}\sum_{i=1}^n \Big\{\wh{\bm \beta}_{\text{pilot}}^\top \wh{\mathbb D}_{ip}\wh{\bm \beta}_{\text{pilot}}  +\wh\sigma^2_{\text{pilot}}\Big\}\wh {\bm U}_i \wh {\bm U}_i^\top, \\
	\text{ and }	\mI_n(\wh \mA) = \Big\{n^{-1}\sum_{i=1}^n \wh {\mathbb D}_{ip} \otimes \Big(\bm W_i\bm W_i^\top\Big)\Big\}^{-1},
\end{gather*}respectively. Here $\wh {\mathbb D}_{ip} = \diag[p(\bm W_i^\top \wh{\bm \alpha}_1)\{1-p(\bm W_i^\top\wh{\bm \alpha}_1)\},\dots, p(\bm W_i^\top \wh {\bm \alpha}_p)\{1-p(\bm W_i^\top\wh{\bm \alpha}_p)\}] \in \mR^{p\times p}$. This leads to a natural estimator for the asymptotic covariance of $\wh{\bm \theta}_{\text{imp}}$ as \beq
\wh \Sigma = \wt \Sigma_u^{-1}\Big\{n^{-1}\wh \Sigma_{uw} \mI_n(\wh \mA) \wh \Sigma_{uw}^\top + N^{-1} \wh \Omega \Big\}\wt\Sigma_u^{-1}.\label{formula unified covariance}\eeq

From Theorem \ref{Theorem regular}, we know that the asymptotic covariance of $\wh{\bm \theta}_{\text{imp}}$ is given as $\cov(\wh{\bm \theta}_{\text{imp}}) =n^{-1}(\Sigma_u - \mW)^{-1}\Sigma_{uw}\mI(\mA)\Sigma_{uw}^\top(\Sigma_u - \mW)^{-1}$. Then, one can easily show that $n\wh \Sigma\stackrel{p}{\to}(\Sigma_u - \mW)^{-1}\Sigma_{uw}\mI(\mA)\Sigma_{uw}^\top(\Sigma_u - \mW)^{-1}$ under the same conditions as in Theorem \ref{Theorem regular}. This seems to be an unsurprising result as $\wh \Sigma$ is constructed accurately to the analytical formula as given by Theorem \ref{Theorem regular}. However, whether it is also a consistent estimator for $\cov(\wh\theta_{\text{imp}})$ under the conditions of Theorems \ref{Theorem imbalanced} and \ref{Theorem high accuracy} is not immediately straightforward. We then have the following theorem.
\begin{theorem}
	(1) Under the same conditions as in Theorem \ref{Theorem imbalanced}, we have $nr_{N,\max}^{-1} \wbD \wh \Sigma \wbD \stackrel{p}{\to} V_{\text{imp}}$;
	(2) Under the same conditions as in Theorem \ref{Theorem high accuracy}, we have $N\wh \Sigma \stackrel{p}{\to} \sigma^2\Sigma_u^{-1}$. \label{Theorem cov}
\end{theorem}
\noindent The detailed proof of Theorem \ref{Theorem cov} is provided in Appendix E in the supplementary material. By Theorem \ref{Theorem cov}, we find that the covariance matrix estimator is still consistent under the same technical conditions as in Theorems \ref{Theorem imbalanced} and \ref{Theorem high accuracy}. Specifically, the convergence rate of $\wh\Sigma$ matches that of $\wh{\bm \theta}_{\text{imp}}$.  In the highly imbalanced case, the estimated covariance of $\wh{\bm\beta}_{\text{imp},j}$ is of order $O_p(n^{-1}r_{N,\max}r_{Nj}^{-2})$ and that of $\wh{\bm \gamma}_{\text{imp}}$ is of order $O_p(n^{-1} r_{N,\max})$, whereas in the highly predictable case, the estimated covariance of $\wh {\bm \theta}_{\text{imp}}$ is of order $O_p(N^{-1})$. Then we can conduct a unified hypothesis test or the unified confidence region to examine the significance of the interested coefficients.

\subsection{A Further Improved Estimator for the Regular Case}

As we have discussed in Theorem \ref{Theorem regular}, the imputed estimator $ \wh{\bm\theta}_{\text{imp}}$ might perform worse than the pilot estimator $\wh{\bm \theta}_{\text{pilot}}$, when $\sigma$ is relative small for the regular case. Then a natural question is: can we find an estimator, which combines the strength of both the imputed and the pilot estimators, so that the new estimator can outperform both the pilot and imputed estimators uniformly? To this end, we consider a weighted estimator as $\wh {\bm \theta}_w = w \wh {\bm \theta}_{\text{pilot}} + (1- w )\wh{\bm \theta}_{\text{imp}}$. Then the pilot and impute estimators are both special cases of $\wh\btheta_w$. Specifically, when $w = 0$, we have $\wh\btheta_0 = \wh\btheta_{\text{pilot}}$. When $w = 1$, we have $\wh\btheta_1 = \wh\btheta_{\text{imp}}$. We next search for the optimal $w \in [0,1]$ for the best asymptotic performance. To this end, we adopt the idea of $A$-optimality \citep{kiefer1959optimum} and define an objective function as $\tr\{\avar(\wh\btheta_w)\} = w^2\tr\{\avar(\wh\btheta_{\text{pilot}})\} + 2w(1-w) \tr\{ \acov(\wh\btheta_{\text{imp}}, \wh \btheta_{\text{imp}})\} + (1-w)^2 \tr\{\avar(\wh\btheta_{\text{imp}})\}$,
where $\avar(\wh\btheta)$ is the asymptotic variance of the estimator $\wh\btheta$ and $\acov(\wh\btheta_1, \wh\btheta_2)$ is the asymptotic covariance matrix between $\wh\btheta_1$ and $\wh\btheta_2$.  By minimizing the loss function, the optimal weight $w^*$ can be calculated as
\[
w^* = \frac{\tr\{\avar(\wh\btheta_{\text{imp}})\} - \tr\{\acov(\wh\btheta_{\text{imp}}, \wh \btheta_{\text{pilot}} )\} }{\tr\{\avar(\wh\btheta_{\text{pilot}})\}  + \tr\{\avar(\wh\btheta_{\text{imp}})\} - 2\tr\{\acov(\wh\btheta_{\text{imp}}, \wh\btheta_{\text{pilot}}) \} },
\]
where $\acov(\wh\btheta_{\text{imp}}, \wh \btheta_{\text{imp}}) = N^{-1}\sigma^2 (\Sigma_u - \mW)^{-1}$. By replacing the unknown parameters by their sample counterparts, we can estimate $w^*$ by
\[
\wh w = \frac{\tr(\wh \Sigma) - N^{-1}\wh \sigma^2_{\text{pilot}}\tr( \wt \Sigma_{u}^{-1})}{n^{-1}\wh\sigma^2_{\text{pilot}} \tr(\wh \Sigma_u^{-1}) + \tr(\wh \Sigma ) - 2N^{-1}\wh \sigma^2_{\text{pilot}}\tr( \wt \Sigma_{u}^{-1})},
\]
where $\wh\Sigma_u = n^{-1}\sum_{i=1}^n \bU_i\bU_i^\top $. We then have the following theorem for the asymptotic efficiency of $\wh\btheta_{\wh w}$.
\begin{theorem}
	Under the same conditions as in Theorem \ref{Theorem regular}, we have $\wh\btheta_{\wh w} - \btheta = \{\bm \zeta_1^*  + (1-w^*)(\Sigma_u - \mW)^{-1}\bm \zeta_2\}\{1+o_p(1)\} $, where $\bm \zeta_1^* = \{w^*\Sigma_u^{-1} + N^{-1}n (1-w^*)(\Sigma_u - \mW)^{-1}\} (n^{-1}\sum_{i=1}^n \bU_i \ve_i) - (1-w^*)(\Sigma_u - \mW)^{-1}\Sigma_{uw}(\wh \mA - \mA)$. \label{Theorem weighted estimator}
\end{theorem}
\noindent The detailed proof of Theorem \ref{Theorem weighted estimator} is provided in Appendix G. By Theorem \ref{Theorem weighted estimator}, we know that $\wh\btheta_{\wh w}$ can be decomposed into two parts. The first part $\bm \zeta_1^*$ is of order  $O_p(n^{-1/2})$ and the second part $(1-w^*)(\Sigma_u - \mW)^{-1}\bm\zeta_2$ is of order $O_p(N^{-1/2})$. Compared with the first part $\bm\zeta_1^*$, the second part is ignorable under the assumption that $n/ N\to0$ as $N \to\infty$. Therefore, the asymptotic covariance of $\wh\btheta_{\wh w}$ is mainly determined by $\bm\zeta_1^*$ with $\avar(\wh\btheta_{\wh w}) \approx n^{-1}\{w^{*2}\sigma^2\Sigma_u^{-1} + (1-w^*)^2 (\Sigma_u - \mW)^{-1}\Sigma_{uw}\mI(\mA)\Sigma_{uw}^\top(\Sigma_u - \mW)^{-1}\}$, which is a weighted covariance matrix of $\wh\btheta_{\text{pilot}}$ and $\wh\btheta_{\text{imp}}$. By specifying $w^* \approx \tr\{(\Sigma_u - \mW)^{-1}\Sigma_{uw}\mI(\mA)\Sigma_{uw}^\top(\Sigma_u - \mW)^{-1}\}/[\tr(\sigma^2\Sigma_u^{-1}) + \tr\{(\Sigma_u - \mW)^{-1}\Sigma_{uw}\mI(\mA)\Sigma_{uw}^\top(\Sigma_u - \mW)^{-1}\} ]$, we then obtain
\begin{eqnarray*}
\avar(\wh\btheta_{\wh w}) &\approx& n^{-1}\frac{\sigma^2\tr\{(\Sigma_u - \mW)^{-1}\Sigma_{uw}\mI(\mA)\Sigma_{uw}^\top(\Sigma_u - \mW)^{-1}\} ^2  \Sigma_u^{-1}}{[\tr(\sigma^2\Sigma_u^{-1}) + \tr\{(\Sigma_u - \mW)^{-1}\Sigma_{uw}\mI(\mA)\Sigma_{uw}^\top(\Sigma_u - \mW)^{-1}\} ]^2} \\
&+& n^{-1}\frac{\tr(\sigma^2 \Sigma_u^{-1})^2 (\Sigma_u - \mW)^{-1}\Sigma_{uw}\mI(\mA)\Sigma_{uw}^\top(\Sigma_u - \mW)^{-1}}{[\tr(\sigma^2\Sigma_u^{-1}) + \tr\{(\Sigma_u - \mW)^{-1}\Sigma_{uw}\mI(\mA)\Sigma_{uw}^\top(\Sigma_u - \mW)^{-1}\} ]^2},
\end{eqnarray*}
whose trace is given by $\tr\{\sigma^2\Sigma_u^{-1}\} \tr\{(\Sigma_u - \mW)^{-1}\Sigma_{uw}\mI(\mA)\Sigma_{uw}^\top(\Sigma_u - \mW)^{-1} \}/ [\tr\{\sigma^2\Sigma_u^{-1}\} + \tr\{(\Sigma_u - \mW)^{-1}\Sigma_{uw}\mI(\mA)\Sigma_{uw}^\top(\Sigma_u - \mW)^{-1} \} ]$. Note that this is a number smaller than $n^{-1}\tr(\sigma^2\Sigma_u^{-1})$ of the pilot estimator $ \wh\btheta_{\text{pilot}}$ and $n^{-1}\tr\{(\Sigma_u - \mW)^{-1}\Sigma_{uw}\mI(\mA)\Sigma_{uw}^\top(\Sigma_u - \mW)^{-1}\}$ for the imputed estimator $\wh\btheta_{\text{imp}}$. In this way, $\wh\btheta_{\wh w}$ is statistically asymptotically more efficient than both $\wh\btheta_{\text{pilot}}$ and $\wh\btheta_{\text{imp}}$.

\section{Simulation Studies}

\subsection{The Finite Sample Performance of $\wh{\theta}_{\text{imp}}$}

To evaluate the finite sample performance of $\wh{\bm \theta}_{\text{imp}}$, we conduct a large number of simulation experiments. We start with the finite sample performance of the three different estimators. They are (1) the \textit{pilot estimator} $\wh{\bm \theta}_{\text{pilot}}$, (2) the \textit{imputed estimator} $\wh{\bm \theta}_{\text{imp}}$, and (3) the OLS estimator $\wh{\bm \theta}_{\text{ols}}$ with fully observed $\bm Z$ for the whole sample without imputation, which is referred to as the \textit{oracle estimator} for convenience. 

The data generation process is described as follows. For the simulation study in Section 3.1 and 3.2, we fix the dimension of $\bm Z_i$ to be $ p = 2$, the dimension of $\bm X_i$ to be $q = 7$ and the dimension of $\bm W_i$ to be $r= 9$. Recall that $\bm W_i = (1,\wt {\bm W}_i^\top)^\top\in\mR^{9}$. Here $\wt {\bm W}_i$'s are independently generated from a multivariate normal distribution with mean $\mathbf 0_{8} \in \mR^8$ to be a zero vector and covariance matrix $\Sigma_w = (\sigma_{w,j_1j_2})_{8\times8}$ where $\sigma_{w,j_1j_2} = 0.25^{|j_1-j_2|}$ \citep{fan2001variable}.  Next, $Z_{ij}$ is generated according to a logistic regression model as $P(Z_{ij} = 1|\bm W_i) = p(\bm W_i^\top\bm \alpha_j) = \exp(\bm W_i^\top\bm \alpha_j) / \{1+\exp(\bm W_i^\top\bm \alpha_j)\}$. 
Write $\bm X_i = (1,\wt {\bm X}_i^\top)^\top$, where $\wt {\bm X}_i$'s are generated from a multivariate normal distribution with mean 1 and covariance matrix $\Sigma_x = (\sigma_{x,ij})_{6\times 6}$ with $\sigma_{x,j_1j_2} = 0.5^{|j_1-j_2|}$. The error terms $\ve_i$ are independently generated as normal random variables with mean 0 and variance $\sigma^2$. Lastly, the responses $Y_i$ are then generated according to $Y_i = \bm Z_i^\top \bm \beta + \bm X_i^\top\bm \gamma + \ve_i$ with $\bm \beta = (3,0)^\top \in \mR^2$ and $\bm \gamma = (1,1.5,0,0,0,2,0)^\top\in\mR^7$. For the first study, we fix the whole sample size to be $N = 200,000$ and the pilot sample size to be $n = 8,000$. For each simulation study, the experiment is replicated a total of $B = 1,000$ times.

Next, we consider the measurements of finite sample performance. Denote $\wh {\bm \theta}^{(b)}$ to be one particular estimator (e.g., the imputed estimator) obtained in the $b$th random replication with $1\le b\le B$. The mean squared error (MSE) is used to evaluate the estimation error for different estimators. Specifically, it is defined as $\MSE(\wh{\bm \theta}^{(b)}_{\mS}) =|\mS|^{-1} \sum_{j\in\mS}(\wh\theta^{(b)}_{j} - \theta_{j})^2$ for $\mS =\{1\},\{2\},\{3,\dots,9\}$ in Example 1 and $\{1,\dots,9\}$ in Example 2, where $\bm \theta_\mS = \{\theta_j:j\in\mS\}$ represents the subvector of $\bm \theta$ with index set $\mS$.  This leads to a total of $B$ $\MSE$ values.

Based on these settings, we consider two examples to demonstrate the imbalance effect and predictability effect for different estimators, which correspond to the two special cases discussed in Section 2.



\begin{figure}[t]
	\centering
	\subfloat[$\wh\beta_1$ when $t = 2$ ]{\includegraphics[width=0.33\columnwidth]{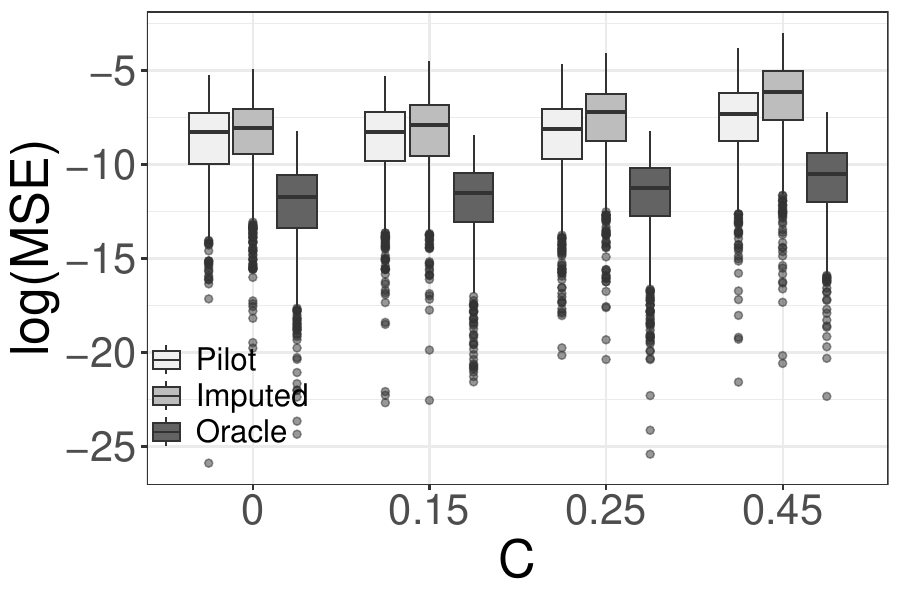}}
	\subfloat[$\wh\beta_2$ when $t = 2$ ]{\includegraphics[width=0.33\columnwidth]{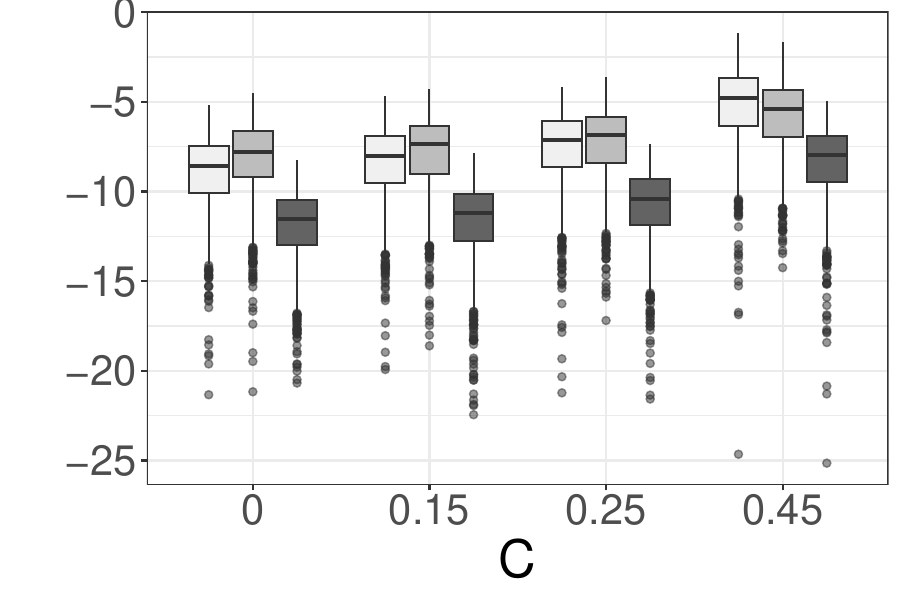}}
	\subfloat[$\wh{\bm \gamma}$ when $t =2$ ]{\includegraphics[width=0.33\columnwidth]{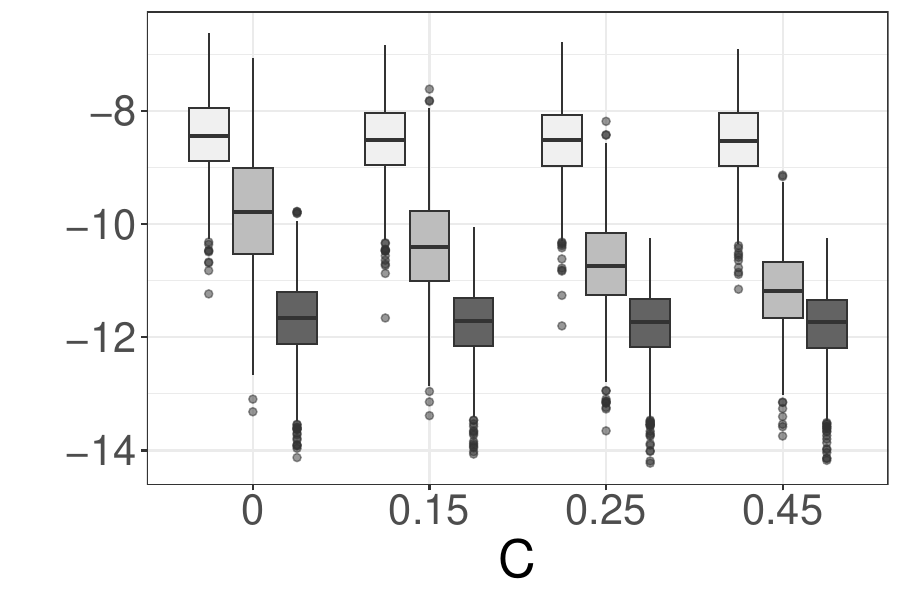}}\\
	\subfloat[$\wh\beta_1$ when $C = 0.25$ ]{\includegraphics[width=0.33\columnwidth]{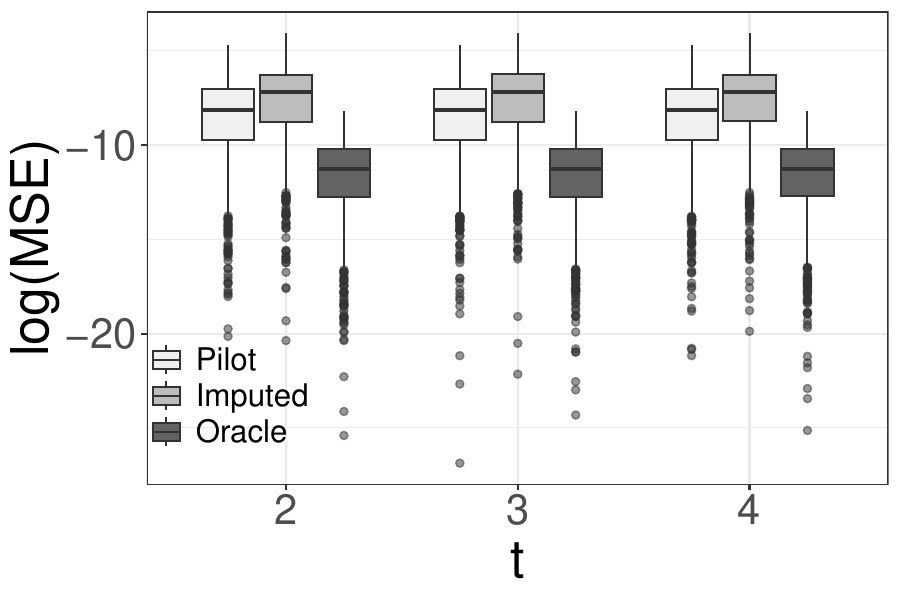}}
	\subfloat[$\wh\beta_2$ when $C = 0.25$ ]{\includegraphics[width=0.33\columnwidth]{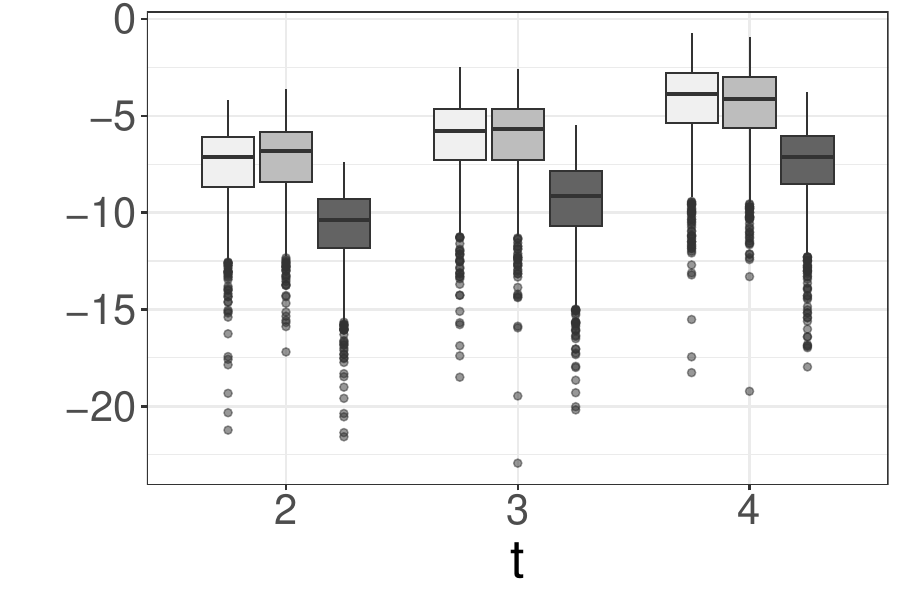}}
	\subfloat[$\wh{\bm \gamma}$ when $C = 0.25$ ]{\includegraphics[width=0.33\columnwidth]{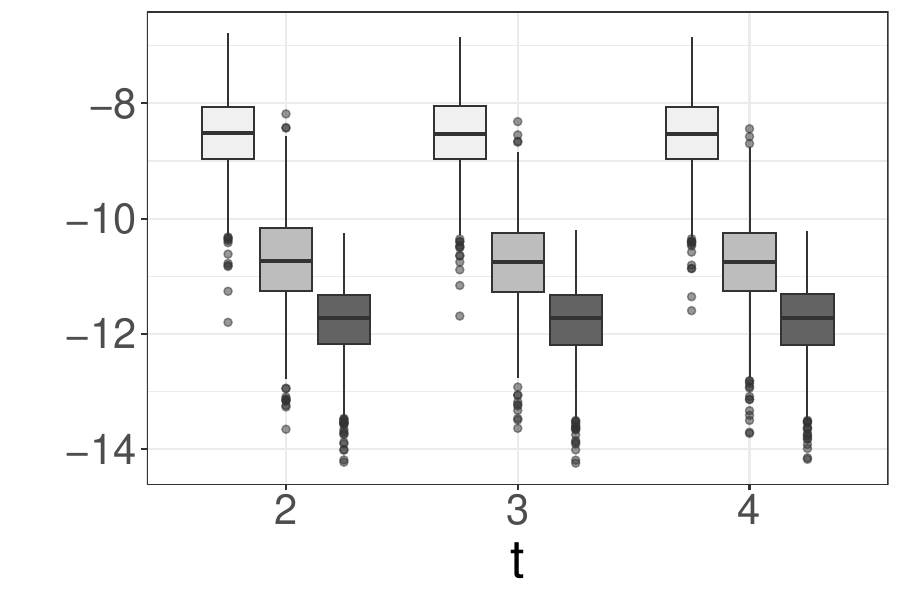}}
	\caption{Boxplots of the log-transformed MSE values of the pilot estimator, the imputed estimator, and the oracle estimator under different settings. }
	\label{Figure imbalanced}
\end{figure}

\noindent  \textbf{Example 1: Imbalance effect. }We first study the imbalance effect with $\sigma = 1$. Consider a logistic regression model with different intercepts. Specifically, set $\bm \alpha_1 = (\alpha_N,3/2,0,0,3/4,0,\\0,-2,0)^\top \in\mR^{9}$ and $\bm \alpha_2 = (t\alpha_N, 1,1,1,-3\sqrt 2/2,1/3,0,0,0)^\top\in\mR^9$ \citep{fan2001variable,fan2009ultrahigh} with $\alpha_N = -C\log(n)$, where $C\in\{0,0.15,0.25,0.45\} $ and $t\in \{2,3,4\}$ are two constants controlling the maximum positive proportion $r_{N,\max}$ and the imbalance between $Z_{i1}$ and $Z_{i2}$, respectively.  It could be calculated that $r_{N,\max} =\max_jr_{Nj} = \exp(\alpha_{N1})= n^{-C}$ and $r_{N2}= n^{-Ct}$. Then an increase in $C$ leads to a decrease in $r_{N,\max}$ and an increase in $t$ results in a decrease in $r_{N2}$. Specifically, here $C=0$ represents the \textit{regular case} with $P(Z_{i1} = 1) \approx 50.0\%$ and $P(Z_{i2} =1 )\approx 50.0\%$; and $C = 0.45$ represents the \textit{highly imbalanced case} with $P(Z_{i1} = 1) \approx 11.0\%$ and $P(Z_{i2} = 1) \le 1\%$. 
The MSE values for $\wh \beta_1$, $\wh \beta_2$ and $\wh {\bm \gamma}$ are then log-transformed and boxplotted in Figure \ref{Figure imbalanced}.

We can draw several conclusions from Figure \ref{Figure imbalanced}. First, the estimation errors of $\wh\beta_{\text{pilot},j}$, $\wh\beta_{\text{imp},j}$, and $\wh\beta_{\text{ols},j}$ increase as $C$ increases with respect to the increase in the medians of the corresponding boxes. This is because the increase in $C$ leads to a decrease in $r_{N,\max}$, which further leads to a slower convergence rate for all the three estimators. This is consistent with the theoretical results in Theorem \ref{Theorem imbalanced}. Second, the estimation errors of $\wh{\bm \gamma}_{\text{pilot}}$ and $\wh{\bm \gamma}_{\text{ols}}$ are relatively stable when $C$ and $r_{N,\max}$ change in the observation that the boxes are located approximately at the same position. By contrast, the estimation error of $\wh{\bm \gamma}_{\text{imp}}$ decreases with the decreasing medians and narrower boxes in the boxplot as $C$ increases and $r_{N,\max}$ decreases. This is consistent with the theoretical claims of Theorem \ref{Theorem imbalanced}. From Theorem \ref{Theorem imbalanced}, we know that $\wh{\bm \gamma}_{\text{pilot}}$ and $\wh{\bm \gamma}_{\text{ols}}$ are not much affected by the imbalance level. By contrast, $\wh{\bm \gamma}_{\text{imp}}$ converges faster as $C$ increases and $r_{N,\max}$ decreases. Third, when $r_{N,\max}$ is fixed and  $r_{N2}$ decreases, the estimation error of $\wh\beta_2$ increases while those of the other estimators remain relatively stable. This is also as expected, as the convergence rate of $\wh\beta_2$ is much affected by $r_{N2}$. 
This explains the worse performance of $\wh\beta_2$. By contrast, the performances of other estimators are not much affected as $r_{N,\max}$ is stable in this case.

\begin{figure}[t]
	\centering
	\subfloat[low noise level case ]{\includegraphics[width=0.45\columnwidth]{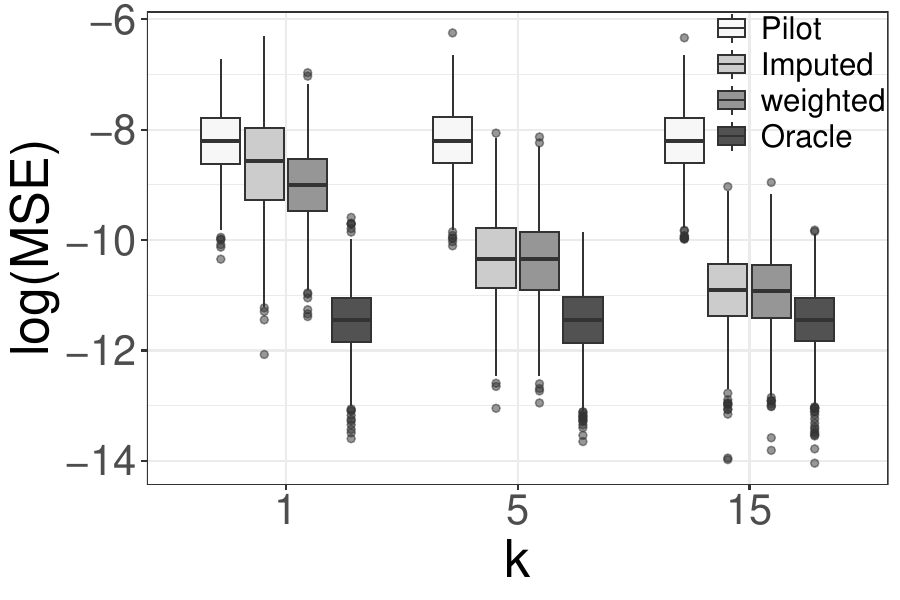}}
	\subfloat[high noise level case ]{\includegraphics[width=0.45\columnwidth]{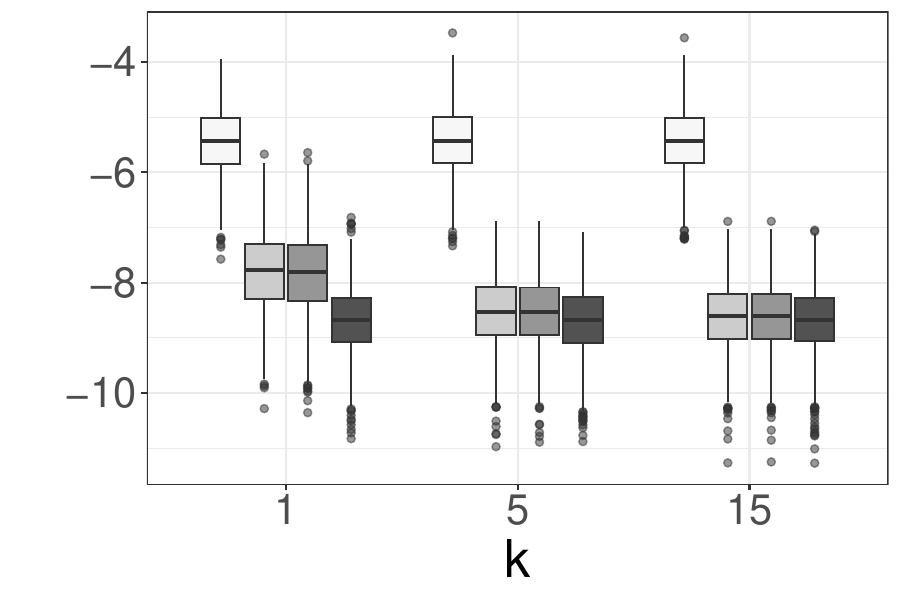}}\\
	\subfloat[regular case]{\includegraphics[width=0.45\columnwidth]{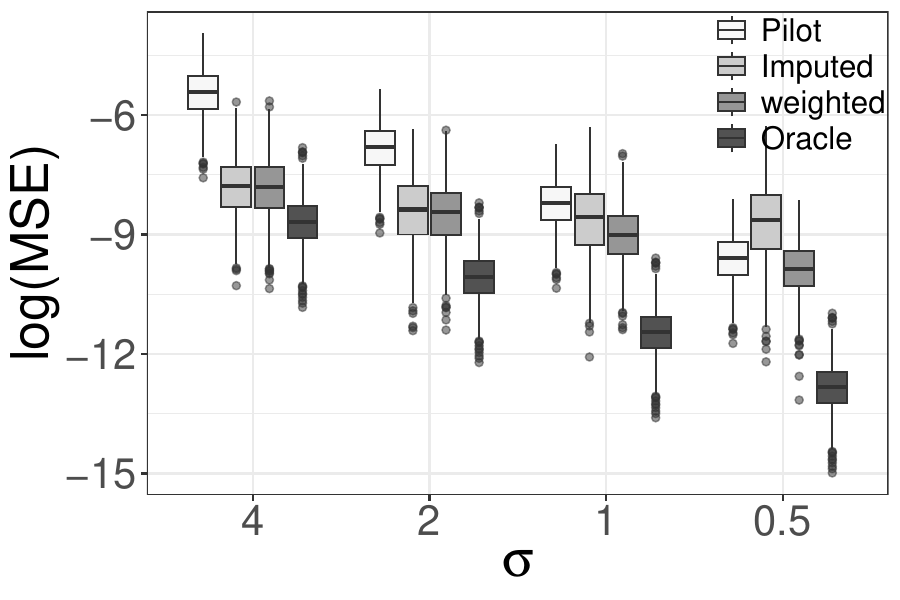}}
	\subfloat[highly predictable case]{\includegraphics[width=0.45\columnwidth]{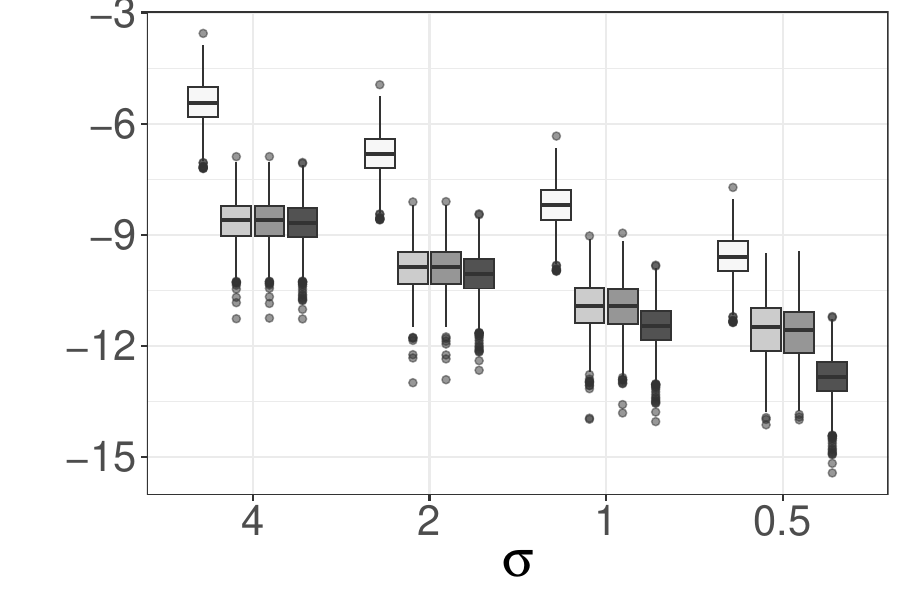}}
	\caption{Boxplots of the log-transformed MSE values of the pilot estimator $\wh{\bm \theta}_{\text{pilot}}$, the imputed estimator $\wh{\bm \theta}_{\text{imp}}$, the oracle estimator $\wh{\bm \theta}_{\text{ols}}$, and the weighted estimator $\wh\btheta_{\wh w}$.}
	\label{Figure high accuracy}
\end{figure}

\noindent \textbf{Example 2: Predictability effect. }We next study examples with different prediction accuracy. Consider a logistic regression model with different coefficient norms. Specifically, set $\bm \alpha_1 = k(0,3/2,0,0,3/4,0,0,-2,0)^\top \in\mR^{9}$ and $\bm \alpha_2 = k(0,1,1,1,-3\sqrt 2/2,1/3,0,0,0)^\top\in\mR^9$, where $k\in\{1,5,15\}$  is some positive constant controlling the size of $\|\bm \alpha_j\|$. Specifically, $\omega_n \approx 0.132,0.032,0.011$ for $k = 1,5,15$.  Then we vary $\sigma \in \{4,2,1,0.5\}$. For different $\sigma$ values, $k = 1$ represents the \textit{regular case} and $k = 15$ represents the \textit{highly predictable case}. For different $k$ values, $\sigma = 4$ represents cases with \textit{high noise level} and $\sigma = 1$ represents cases with \textit{low noise level}. In addition to the aforementioned three estimators, we also consider the weighted estimator $\wh\btheta_{\wh w}$ in this example. 
The \MSE\, values for $\wh{\bm \theta}$ are then log-transformed and boxplotted in Figure \ref{Figure high accuracy}.

We can obtain the following conclusions from Figure \ref{Figure high accuracy}. First, as $k$ increases and $\omega_n$ decreases, we find that the $\log(\MSE)$ of $\wh{\bm \theta}_{\text{imp}}$ approaches that of the oracle estimator $\wh{\bm \theta}_{\text{ols}}$ with a smaller difference between the two estimators in the median values of $\log(\MSE)$ in both low and high noise level cases. This finding is consistent with the theoretical claim of Theorem \ref{Theorem high accuracy}. 
As shown in Appendix E, we have $\bm \zeta_1 = O_p(\omega_n^{1/2}n^{-1/2})$. In this way, the decrease in $\omega_n$ leads to the decrease in $\bm \zeta_1$. Consequently, $\wh{\bm \theta}_{\text{imp}}$ becomes closer to $\bm \zeta_2$, whose asymptotic variance is approaching the same as that of $\wh{\bm \theta}_{\text{ols}}$. Second, for both the regular and highly predictable cases, the noise level (as controlled by $\sigma$) does affect the finite sample performance of the pilot sample and oracle estimators significantly. For the cases with high noise level ($\sigma = 4$ in the simulation), the performances of the imputed estimators are fairly close to those of the oracle estimator $\wh{\bm \theta}_{\text{ols}}$. This is as expected as the estimation error from the imputation model is relatively small compared with the regression error in the substantive model. Whereas for the case with low noise level ($\sigma = 1$ in the simulation), $\wh{\bm \theta}_{\text{imp}}$ performs poorly and could be even worse than $\wh{\bm \theta}_{\text{pilot}}$ in the sense of the MSE in the regular case. This is also as expected, because by Theorem \ref{Theorem regular} the regression error will be dominated by the estimation error from the pilot sample size based logistic regression. By contrast, $\wh{\bm \theta}_{\text{imp}}$ performs much better than $\wh {\bm \theta}_{\text{pilot}}$ in the highly predictable case even though the noise low level is low. This is also expected, because by Theorem \ref{Theorem high accuracy} the convergence rate of $\wh{\bm \theta}_{\text{imp}}$ is faster than that of $\wh{\bm \theta}_{\text{pilot}}$. All these numerical findings corroborate the theoretical claims of Theorem \ref{Theorem high accuracy} very well. Lastly, the values of $\log(\MSE)$ for the weighted estimator $\wh\btheta_{\wh w}$ are always smaller than those of $\wh\btheta_{\text{imp}}$ and $\wh\btheta_{\text{pilot}}$ in all cases, which is consistent with the results in Theorem \ref{Theorem weighted estimator}.

\subsection{Finite Sample Performance of the Unified Covariance Estimator}

Next, we verify the finite sample performance of the unified covariance estimator $\wh \Sigma$. In this section, we consider three different cases, which are listed as follows.

\textbf{Case 1. Regular case.} We consider $C =0$ in Example 1.

\textbf{Case 2. Highly imbalanced case. }We set $C = 0.45$ and $t = 2$ in Example 1.

\textbf{Case 3. Highly predictable case. }We let $k = 15$ and $\sigma = 1$ in Example 2.

For the sample sizes, let $(n,N) \in \{(6000,140000), (8000,200000)\}$. For each $\wh\theta_j$, the true standard error is estimated by $\SE_j = \{(B-1)^{-1}\sum_{b=1}^B(\wh\theta_j^{(b)} - \bar \theta_j)^2\}^{1/2}$, where $ \bar\theta_j = B^{-1}\sum_{b=1}^B\wh\theta_j^{(b)} $. Let $\wh {\SE}_j^{(b)}$ be the estimated standard error of $\wh\theta_j^{(b)}$ using the unified covariance estimator \eqref{formula unified covariance} from the $b$th replication. Define $\wh {\SE}_j = B^{-1}\sum_{b=1}^B \wh {\SE}_j^{(b)}$. We also compute the empirical coverage probability of the 95\% confidence interval, which is given as $\CP(\wh\theta_j) = B^{-1}\sum_{b=1}^B I\big(\theta_j \in [\wh\theta_j^{(b)} - 1.96 \wh {\SE}_j^{(b)}, \wh\theta_j^{(b)} + 1.96 \wh {\SE}_j^{(b)}]\big)$. We consider two types of estimators here. The first type of estimators are $\wh {\SE}_j^{\text{obs}}$ and $\CP^{\text{obs}}$. They are, respectively, the standard error estimate obtained by treating the imputed values as if they were the true values and the coverage probability of the resulting confidence interval. The second type of estimators are $\wh {\SE}_j^{\text{imp}}$ and $\CP^{\text{imp}}$. They are, respectively, the standard error estimate obtained by using the unified covariance estimator \eqref{formula unified covariance} and the coverage probability of the resulting confidence interval. The simulation results for different cases are presented in Table \ref{Table cov est}. 

\begin{table}[t]
\centering
\caption{Simulation results for covariance estimators under different cases  and different sample sizes. The $\SE$ and $\wh \SE$ values are multiplied by $10^{-2}$. }
\begin{tabular}{cccccccccc}
	\hline
	\hline
	Estimators                 & $\beta_1$ & $\beta_2$  &  $\gamma_1$ & $\gamma_2$ & $\gamma_3$ & $\gamma_4$ & $\gamma_5$ & $\gamma_6$ & $\gamma_7$ \\
	\hline
	\multicolumn{10}{c}{Regular case $n=6000, N = 140000$ } \\
	$\SE_j$     &2.867&	3.614&	2.723&	0.434&	0.510&	0.506&	0.517&	0.492&	0.458
	\\
 $\wh {\SE}_j^{\text{obs}}$ &  0.761&0.742&	0.730&	 0.309&	0.345&	0.345&	0.345&	 0.345&	 0.309
	\\
$\CP^{\text{obs}}(\%)$ & 40.1&	29.9&	38.2&	82.8&	80.1&81.4&	80.4&	 83.8&81.4
	\\
	$\wh {\SE}_j^{\text{imp}}$ & 3.134&	3.937&	2.915&	0.461&	0.516&	0.516&	0.516&	0.516&	0.461
	\\
	$\CP^{\text{imp}}(\%)$ &96.0&	96.9&	96.9&	96.2&	96.1&	96.6&	94.4&	95.7&	94.9
	\\

	\multicolumn{10}{c}{Regular case $n=8000,N = 200000$ }     \\
	$\SE_j$     &2.580&	3.121&	2.374&	0.373&	0.437&	0.431&	0.428&	0.427&	0.367
	\\
$\wh {\SE}_j^{\text{obs}}$ & 0.637& 0.622 & 0.611 & 0.258 & 0.289 & 0.289 & 0.289 & 0.289 & 0.268
	\\
$\CP^{\text{obs}}(\%)$ & 36.3 & 29.8 & 36.5 & 82.7 & 80.8 & 80.2& 82.4 &82.1 & 84.2
	\\
	$\wh {\SE}_j^{\text{imp}}$ & 2.704&	3.403&	2.513&	0.385&	0.430&	0.430&	0.430&	0.431&	0.385
	\\
	$\CP^{\text{imp}}(\%)$ &95.3&	97.0&	96.8&	95.8&	94.0&	95.3&	94.6&	95.2&	96.1
	\\
	\hline
	\multicolumn{10}{c}{Highly imbalanced case $n = 6000,N = 140000$ }                                                                                                                                  \\
	$\SE_j$    & 7.142&	11.558&	0.926&	0.379&	0.412&	0.427&	0.423&	0.414&	0.378\\
	
$\wh {\SE}_j^{\text{obs}}$ & 1.423 & 6.123 & 0.533 & 0.309 & 0.345 & 0.345 & 0.345 & 0.345 & 0.309 \\
$\CP^{\text{obs}}(\%)$ &31.9 & 70.8 & 72.5 & 89.0 & 89.3 & 89.2 & 88.5 & 89.3 & 89.0
	\\
	$\wh {\SE}_j^{\text{imp}}$ &8.176&	14.010&	0.960&	0.384&	0.429&	0.429&	0.429&	0.429	&0.384
	\\
	$\CP^{\text{imp}}(\%)$ &97.0&	98.2&	96.1&	95.9&	96.3&	95.2&	95.2&	96.2&	96.1
	\\
	\multicolumn{10}{c}{Highly imbalanced case $n = 8000,N = 200000$}  \\
	$\SE_j$    & 6.757&	10.312&	0.759&	0.319&	0.345&	0.348&	0.341&	0.352&	0.310
	\\
$\wh {\SE}_j^{\text{obs}}$ & 1.242 & 5.847 & 0.446 & 0.258 & 0.289 & 0.289 & 0.289 & 0.289 & 0.258 \\
$\CP^{\text{obs}}(\%)$ & 27.1& 75.1 & 75.3 & 89.2 & 90.1 & 89.5 & 89.7 & 88.9 &89.6
	\\
	$\wh {\SE}_j^{\text{imp}}$ &7.411&	12.995&	0.797&	0.318&	0.355&	0.355&	0.355&	0.355&	0.317
	\\
	$\CP^{\text{imp}}(\%)$ &96.3&98.5&	94.9&	95.6&	95.3&	95.1&	96.2&	95.1&	95.5
	\\
	\hline
	\multicolumn{10}{c}{Highly predictable case $n = 6000,N = 140000$}  \\
	$\SE_j$    & 0.719&	0.859&	0.848&	0.316	&0.354	&0.361	&0.362&	0.354&	0.328
	\\
$\wh {\SE}_j^{\text{obs}}$ &0.546 & 0.545 & 0.637 & 0.309 & 0.345 & 0.345 & 0.345 & 0.345 & 0.309 \\
$\CP^{\text{obs}}(\%)$ & 83.8 & 77.8 & 85.9 & 94.6 & 95.0 & 93.7 & 94.1 & 93.6 &93.1
	\\
	$\wh {\SE}_j^{\text{imp}}$ &0.720	&0.876&	0.873&	0.330&	0.369&	0.369	&0.369&	0.369&	0.330
	\\
	$\CP^{\text{imp}}(\%)$ &93.0&	94.9&	95.2&	96.1&	96.7&	95.7&	95.7&	95.5&	94.8
	\\
	\multicolumn{10}{c}{Highly predictable case $n = 8000,N = 200000$ }                                                                                                                                  \\
	$\SE_j$    & 0.600&	0.757&	0.746&	0.270&	0.304&	0.305&	0.291&	0.299&	0.270
	\\
$\wh {\SE}_j^{\text{obs}}$ & 0.457 & 0.457 & 0.533 & 0.258 & 0.289 & 0.289 & 0.289 & 0.289 & 0.258 \\
$\CP^{\text{obs}}(\%)$ & 82.4 &75.0 & 81.9 & 95.2 & 93.8 & 93.7 & 94.5 &93.5 & 93.2
	\\
	$\wh {\SE}_j^{\text{imp}}$ &0.611&	0.747&	0.739&	0.275&	0.307&	0.307&	0.307&	0.308&	0.275
	\\
	$\CP^{\text{imp}}(\%)$ &93.5&	94.5&	94.1	&96.7	&95.8	&95.4&	95.9&	94.8&	95.6\\
	%
	%
	%
	%
	%
	%
	\hline
\end{tabular}
\label{Table cov est}
\end{table}

From Table \ref{Table cov est}, we find that the unified covariance estimators work fairly well for different cases. First, for all the settings, the values of $\wh \SE^{\text{imp}}$ are quite close to the true $\SE$. In contrast, the values of $\wh \SE^{\text{obs}}$ are very different from those of the true $\SE$. Meanwhile, as $n$ and $N$ increase, the differences between $\wh \SE^{\text{imp}}$ and $\SE$  shrink towards 0. This result suggests that $\wh \SE^{\text{imp}}$ is a consistent estimator for the true $\SE$. This corroborates the theoretical claims of Theorem \ref{Theorem cov} very well. As a consequence, we find that the empirical coverage probability (i.e., $\CP$) of the confidence interval constructed by using the unified covariance estimator is fairly close to their nominal level of 95\%. In contrast, those of $\CP^{\text{obs}}$ are far from 95\%.  This further confirms that $\wh \SE_j^{\text{obs}}$ is not a consistent estimator for the true $\SE$ at all. 
Second, compared with the regular case, the $\SE$ and $\wh\SE$ values of $\wh\beta_{\text{imp},1}$ and $\wh\beta_{\text{imp},2}$ increase, while  those of $\wh{\bm \gamma}$ decrease slightly in the highly imbalanced case. This is also consistent with the theoretical claims in Theorems \ref{Theorem imbalanced} and \ref{Theorem cov}. By Theorems \ref{Theorem imbalanced} and \ref{Theorem cov}, we know that the variance and the estimated variance of $\wh\beta_{\text{imp},j}$ are all of order $O(n^{-1}r_{N,\max}^{-1})$ and those of $\wh{\bm \gamma}_{\text{imp}}$ are of order $O(n^{-1}r_{N,\max})$. Consequently, the increase in the imbalance level will result in a larger variance in $\wh{\bm \beta}_{\text{imp}}$ and a smaller variance in $\wh{\bm \gamma}_{\text{imp}}$.
Third, compared with the regular case, the $\SE$ and $\wh \SE$ values of $\wh{\bm \beta}_{\text{imp}}$ and $\wh {\bm \gamma}_{\text{imp}}$ also decrease for the highly predictable case. This is consistent with the theoretical findings in Theorems \ref{Theorem high accuracy} and \ref{Theorem cov}. As shown in Theorems \ref{Theorem high accuracy} and \ref{Theorem cov}, in the highly predictable case, the variance and estimated variance of $\wh{\bm \theta}_{\text{imp}}$ are of order $O(N^{-1})$. Under the assumption $n/N\to0$, the performance of $\wh{\bm \theta}_{\text{imp}}$ in the highly predictable case will become much better than the regular case. All the simulation results demonstrate the empirical effectiveness of the unified covariance estimator.

\subsection{Imputation by Deep Neural Networks}
 In this subsection, we present an experiment with the binary variables $Z_{ij}$'s imputed by some sophisticated deep neural network models. More specifically, if we extract the $\bW_i$-feature from a pre-trained neural network model without fine tuning, then our method for $\bW_i$ and $\bZ_i$ is a standard logistic regression. It is remarkable that fine tuning refers to making small adjustments to the parameter estimates of the pre-trained model by the additional information provided by the target dataset, so that the resulting statistical performance can be further improved; see \cite{oquab2014learning} and \cite{rebuffi2017learning} for a more detailed discussion. However, if we allow the parameter in the neural network for $\bW_i$-feature extraction to be fine tuned based on our data, our method becomes a deep neural network model immediately. We consider two different simulation models here.

  \textbf{Model 1.} The true regression model between $\bW_i$ and $\bZ_i$ is assumed to be a simple logistic regression model. More specifically, the generation of $\bW_i$ and $\bZ_i$ is the same as that in Example 3.2 with $k = 1$.

\textbf{Model 2.}  We assume to model the relationship between $\bW_i$ and $\bZ_i$ by a highly nonlinear regression model.  First, $\bW_i= (1, \wt \bW_i^\top )^\top \in \mR^9$ is generated in the same way as before. Next we generate $Z_{i1}$ and $Z_{i2}$ by $P(Z_{ij} = 1|\bW_i ) = p\{f_j(\bW_i)\}$ where  $f_1(\bW_i) =   \{2W_{i1}W_{i2} -3 \sin\{(W_{i3}+ W_{i4})(W_{i5} - 0.2)\} + \cos\{2(W_{i7} - 2.5)W_{i8}\}$ and $f_2(\bW_i) = 2W_{i1}W_{i2} + 2.4W_{i3}W_{i4}W_{i5} -  4\sin(W_{i6}W_{i7}) +2W_{i8} $. The data generation of $\bX_i$ and $\ve_i$ is the same as Section 3.1. Once $\bZ_i$ and $\bX_i$ are given, we have $Y_i$ generated in the same way as Example 3.2 with $\sigma = 1$.

\begin{figure}[t]
 	\centering
 	\subfloat[Logistic regression model case ]{\includegraphics[width=0.45\columnwidth]{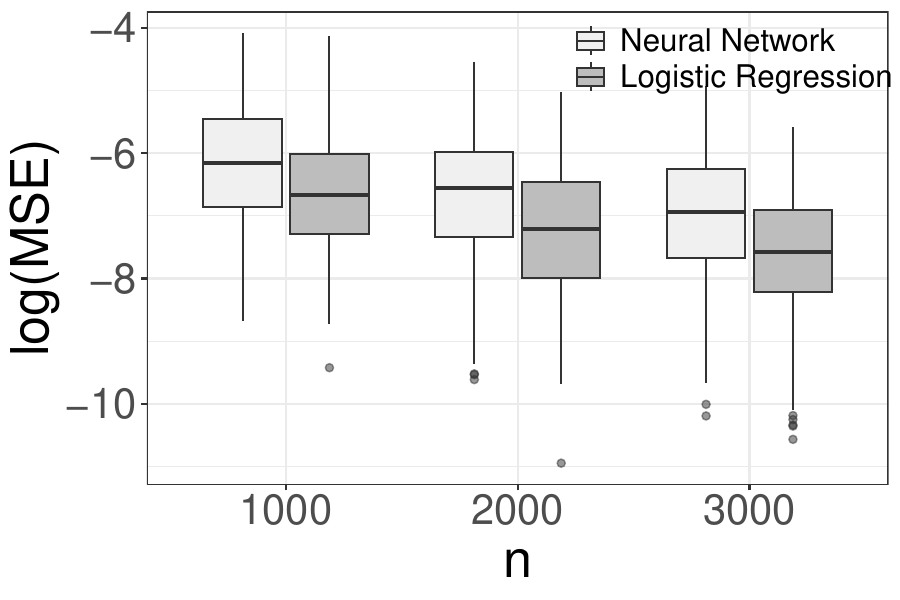}}
 	\subfloat[Highly nonlinear model case ]{\includegraphics[width=0.45\columnwidth]{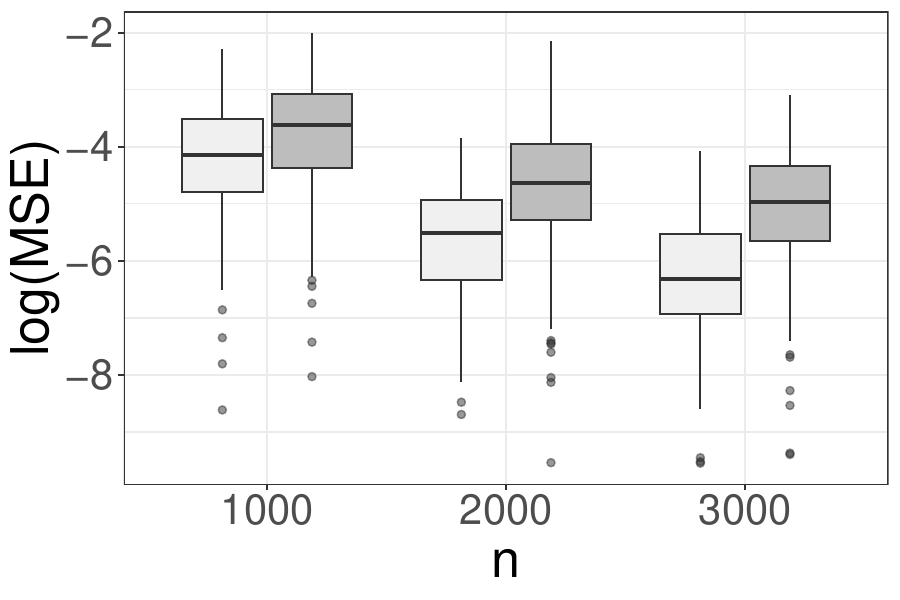}}
 	\caption{Boxplots of the log-transformed MSE values of the {imputed estimator $\wh\btheta_{\text{imp}}$} under different settings.}
 	\label{Figure neural network}
 \end{figure}

We next consider estimating the regression relationship between $\bW_i$ and $\bZ_i$ by a deep neural network regardless of the fact whether the true regression relationship is a single logistic regression model or a highly nonlinear one. Consider a full connected neural network with a total of 4-layers. The GELU function is used for each hidden neural activation \citep{hendrycks2016gaussian}. The width of each hidden layer is fixed to be 16, 32 and 16, respectively. To avoid overfitting, a dropout layer with a 50\% dropout rate is used between the second and third layers for training \citep{srivastava2014dropout}. This leads to a total of 1,304 parameters. We next vary the sample size to be $(n,N) \in \{(1000,25000), (2000,60000),(3000,100000)\}$. For each simulation example, the experiment is replicated for a total of $B = 200$ times. This leads to a total of 200 MSE values. They are then boxplotted in Figure \ref{Figure neural network}.  For comparison purpose, the binary covariates imputed by the logistic regression model are also reported.

 From Figure \ref{Figure neural network}, we find that as $(n,N)$ increases, the $\log(\MSE)$ of $\wh\btheta_{\text{imp}}$ becomes smaller for both models and both imputation methods. In the meanwhile, the $\log(\MSE)$ of $\wh\btheta_{\text{imp}}$ imputed by logistic regression model is smaller than that of $\wh\btheta_{\text{imp}}$ imputed by neural networks, if the true model is logistic regression model. In contrast, the estimator $\wh\btheta_{\text{imp}}$ imputed by neural network outperforms the one with missing features imputed by the logistic regression model, if the relationship between $\bW_i$ and $\bZ_i$ is highly non-linear.

\section{Emotion Recognition Analysis}

We present here a real data example to demonstrate the practical usefulness of the proposed procedure. It is about the aforementioned audio record dataset. The raw data contains a total of 3,262 live streaming video records. These video records are generated by automobile dealers promoting their automobile products on the DouYin platform in China. Each video record lasts for about 30 to 150 minutes, which leads to approximately 253,459 minutes video records in total. These video records are then cut into a total of 2,744,173 video clips with each clip lasting 5 seconds. Each clip is then treated as one observation leading to a large sample size of $N = 2,744,173$. For each observation, the response of interest $Y_i$ is the increment of the number of likes during the live streaming clip in log scale. 
Intuitively, the number of likes reflects the popularity and quantities of the live streaming and thus is of great practical importance.
Therefore, it makes sense to examine the factors influencing this response.

An important factor is the live streamer's emotional state. Intuitively, if the live streamer demonstrates strong passion and confidence, the audience are more likely to generate likes. However, this intuition has never been empirically verified by rigorous statistical analysis. To empirically test this hypothesis, we might need to code each observation (i.e., each 5 second video clip) for two binary indicators $Z_{i1}$ and $Z_{i2}$ to describe the streamer's emotional state. Here $Z_{i1}$ is the \textit{valance} of the live streamer's emotion with $Z_{i1} = 1$ representing positive and $Z_{i1} = 0$ otherwise. Moreover, $Z_{i2}$ is the \textit{arousal} of the live streamer's emotion with $Z_{i2} = 1$ representing a strong emotion and $Z_{i2}=0$ otherwise \citep{russell1980circumplex}. Unfortunately, this is a task relying on human effort (i.e., field experts). Therefore, it will be extremely expensive to process the whole sample $N = 2,744,173$. To alleviate the data labeling cost, we then have to rely on the imputation method developed in this work with a pilot sample of size $n =6,568$ according to our limited budget. The emotional state of each pilot sample is then manually coded by field experts. Next, for the rest uncoded observations, the missing emotional state is imputed with an $r$-dimensional feature vector $\bW_i= (1,\wt {\bm W}_i^\top)^\top\in\mR^{r}$ with $r = 7$. The feature vector $\bW_i$ is constructed using the Mel spectrogram \citep{rabiner2010theory} together with a multi-task trained VGG16 model with the weights pretrained on ImageNet \citep{lecun1989handwritten,caruana1997multitask,simonyan2014very}. Then the first 6 principal components $\wt {\bm W}_i$ are extracted from the output of the last layer of the neural network. 
Thereafter, a logistic regression model from $\bm W_i$ to each $Z_{ij}$ can be constructed. This leads to an imputation model with an out-of-sample AUC 64.27\% for $Z_{i1}$ and 81.74\% for imputing $Z_{i2}$. Moreover, the cases with $Z_{i1}=1$ accounts for approximately 5.36\% in the pilot sample whereas $Z_{i2}=1$ accounts for about 17.44\%. Therefore, this particular dataset seems to fall into the class of regular cases.


Other than emotional state, we also collect for each observation a set of three explanatory variables as $\bm X_i$, which are (1) the number of total followers before the live streaming; (2) the cumulative likes before this period; (3) the number of danmaku characters, see \cite{wang2022impact} and \cite{zhou2022study}. All the covariates are log-transformed and plus 1 before the transformation to avoid zero values. 
Subsequently, our method is applied and the detailed estimation results are given in Table \ref{Table real result}. For comparison purposes, those of the pilot estimate are also presented. 
We find that the estimated standard errors of both $\wh{\bm \beta}_{\text{pilot}}$ and $\wh{\bm \gamma}_{\text{pilot}}$ are significantly larger than those of $\wh{\bm \beta}_{\text{imp}}$ and $\wh{\bm \gamma}_{\text{imp}}$, respectively. This suggests that a statistically more accurate estimator can be obtained using the imputation method. Both models suggest that both positive and strong emotions have a significantly positive effect on the number of likes. In addition to that, the number of followers, cumulative likes, and danmaku characters all have significantly positive effects on the current number of likes. All results are consistent with our expectations, but we have provided quantitative results with statistical significance. To gain some intuitive understanding about the practical significance of both $\bZ_i$-features, we compute the standard deviation of the response, which is given by 0.571. We then use this as a reference. Note that the estimated coefficient of $Z_{i1}$ and $Z_{i2}$ based on imputed data are given by 0.040 and 0.038, respectively. As a result, the relative effect size of the total effect due to emotion measured by the ratio between the coefficients and the standard deviation of the response is given by $(0.040+0.038) / 0.571 = 13.7\%$.
\begin{table}[t]
	\centering
	\caption{The estimation results of the audio record dataset. The regression coefficients with $p$-value less than 0.05 and larger than 0.01 are highlighted with **. The regression coefficients with $p$-value less than 0.01 are highlighted with ***. }
	\begin{tabular}{c|rr|rr}
		\hline
		\hline
		& \multicolumn{2}{c|}{Pilot}           & \multicolumn{2}{c}{Imputed} \\
		& Estimate & S.E.             & Estimate & S.E.        \\
		\hline
		Intercept &  $-0.323^{***}$  & 0.0527 & $-0.348^{***}$    & 0.0026  \\
		$X_1 = $ (log) followers   & $0.018^{***}$     & 0.0048  & $0.021^{***} $   & 0.0002\\
		$X_2=$ (log) cum. likes     & $0.081^{***} $   & 0.0042 & $0.079^{***} $   & 0.0002\\
		$X_3 = $ (log) danmaku char.   & $0.067^{***}$    & 0.0062   &  $0.065^{***}$   & 0.0003  \\
		\hline
		Valance $Z_1=1$  & $0.014{\quad \, \, \, }$    & 0.0295    & $0.040^{***}$   & 0.0082  \\
		Arousal	$Z_2 = 1$   & $0.043^{**\ \,}$     & 0.0183    & $0.038^{***}$   & 0.0025 \\
		\hline
	\end{tabular}
	\label{Table real result}
\end{table}

We next randomly split the datasets into two parts with 80\% of the observations for training and the rest for testing. Then we estimate the coefficient on the training data and compute the RMSE on the testing one. We compare two models. The first model is the linear regression model \eqref{linear regression} with missing $\bZ_i$-feature imputed by $\wh \bZ_i$. The second model is almost the same as the first one but with the feature vector $\bW_i$ also included. We find that the out-of-sample prediction accuracy as measured by RMSE is almost unchanged with 0.5436 for the first model and 0.5434 for the second one. This suggests that $\bW_i$ might have little extra contribution for improving prediction accuracy.

\section{Concluding Remarks}

In this article, we study the theoretical properties of imputed estimator of linear regression with imputed binary covariates. Rigorous asymptotic theory is established for the regular case. Different special cases (i.e., the highly imbalanced and highly predictable cases) are also investigated and theoretically discussed. A unified covariance matrix estimator is also developed for statistical inference. Extensive numerical studies are presented to demonstrate the proposed method. 
To conclude the article, we discuss here a number of interesting topics for future research. First, this study assumes a fixed-dimensional feature. Then, how to handle high dimensional features becomes an interesting problem. Second, this study uses a standard linear regression model. This leaves many other important models (e.g., generalized linear models) open for study. Third, for the highly imbalanced case, the convergence rate of the imputed estimator is unfortunately slower than that of the pilot estimator. Then how to obtain a better estimator remains an important and challenging problem. Lastly, our working imputation model is also related to a single index model as $Y_i = \sum_{j=1}^{p} \beta_{j} \sigma(\bW_{i}^{\top} \alpha_{j})+\bX_{i}^{\top} \beta+\ve_{i}$, where $\sigma(\cdot)$ is the sigmoid function. If this is assumed to be true population model, how to conduct imputation and then valid statistical inference is also an interesting problem for further study.
\begin{supplement}%
\stitle{Technical proofs and additional numerical results}
\sdescription{The Supplemental Material contains some useful lemmas, the technical proofs of Theorem 1-5, the verification of the results in Section 2.4 and some additional numerical results. }
\end{supplement}
%
%
\bibliographystyle{imsart-nameyear}
\bibliography{reference}

\end{document}